\documentclass[twocolumn,aps,prl,showpacs,preprintnumbers,amsmath,amssymb,floatfix]{revtex4-1}
\usepackage[utf8]{inputenc}
\usepackage{graphicx}
\usepackage{amsfonts}
\usepackage{CJK}
\usepackage{bm}
\usepackage{color}
\usepackage{xcolor}
\usepackage{appendix}
\usepackage{mathrsfs}
\usepackage[colorlinks = true,linkcolor = blue,urlcolor = blue,citecolor = blue,anchorcolor = blue]{hyperref}

\makeatletter

\newcommand{\lyxmathsym}[1]{\ifmmode\begingroup\def\b@ld{bold}
  \text{\ifx\math@version\b@ld\bfseries\fi#1}\endgroup\else#1\fi}

\makeatother

\begin{document}
\title{Geometric Bound for Trade-off Relation in Quantum Tricycle}
\author{Shihao Xia}
\thanks{These authors contributed equally to this work.}
\author{Jingyi Chen}
\thanks{These authors contributed equally to this work.}
\author{Jincan Chen}
\author{Shanhe Su}
\email{sushanhe@xmu.edu.cn}

\affiliation{Department of Physics, Xiamen University, Xiamen, 361005, People's
Republic of China }
\date{\today}
\begin{abstract}
We establish a finite-time quantum tricycle driven by an external
field and investigate its thermodynamic performance in the slow-driving
regime. By developing a perturbative expansion of heat with respect
to operation time, we capture the dynamics of heat exchange processes
beyond the quasistatic limit. Within a geometric framework, we derive
fundamental bounds on trade-offs between the cooling rate, coefficient
of performance, and dissipation, governed by the thermodynamic length
and trajectory geometry in control space. Our findings unveil intrinsic
limits to the performance of quantum thermal machines and highlight
the role of geometry in shaping finite-time thermodynamics. This work
advances the fundamental understanding of quantum thermodynamic processes
and offers guiding principles for the design of next-generation quantum
technologies.
\end{abstract}
\maketitle
\textit{Introduction.}---Recent developments in quantum thermodynamics
have pushed the boundaries of thermal machine research into the microscopic
domain, where quantum coherence, entanglement, and finite-time effects
become essential to performance and functionality \cite{Broeck2005,Curzon1975,Salamon1980,Sekimoto1997,Esposito2009,Tu2008,Izumida2009,Izumida2010,Miller2019}.
Among various configurations, systems coupled to three heat reservoirs
have garnered significant attention as minimal yet versatile models
that encapsulate key aspects of quantum thermodynamic processes and
energy conversion \cite{Kosloff2013,Guo2019,Yan1989,Chen1989,Chen1989PRA}.
These devices, often termed quantum tricycles, extend conventional
two-reservoir setups by introducing a third thermal bath, thereby
enabling enhanced control, richer dynamics, and novel modes of operation.

While classical thermodynamics sets fundamental limits on the performance
of quasi-static cycles, practical machines must operate in finite
time to generate finite cooling rate or power \cite{Schmiedl2007,Abiuso2020,Scandi2018,Esposito2010,Tomas2012,Wang2012,Gonzalez-Ayala2018,Holubec2020,Ye2021}.This
necessity inevitably gives rise to trade-offs between key performance
indicators, including the available energy, irreversible entropy generation,
and operation time \cite{Tomas2013,Miller2021,Pietzonka2018,key-57,key-60,key-66,key-70,key-73,key-76,key-79,key-80}.
Understanding these trade-offs is crucial for optimizing the performance
of quantum thermal machines and has spurred significant theoretical
advances in finite-time thermodynamics.

In the slow-driving regime of quantum systems, perturbative approaches
enable a systematic analysis of heat exchange dynamics, uncovering
a direct relationship between the dissipation and the duration of
the thermodynamic cycle \cite{Cavina2017,Breuer2002,Rivas2012}. Such
analyses have led to universal trade-off relations that balance the
energy conversion efficiency against power or cooling rate. In particular,
our work reveals a fundamental thermodynamic constraint for a quantum
refrigerator operating with three heat reservoirs:

\begin{equation}
R\left(\frac{\varepsilon_{r}}{\varepsilon}-1\right)\geq\frac{\bar{\mathcal{\mathscr{L}}^{2}}}{\tau}\label{eq:bound}
\end{equation}
where the cooling rate $R$ and coefficient of performance (COP) $\varepsilon$
is constrained by the driving speed $\tau$ and the reduced thermodynamic
length $\bar{\mathcal{\mathscr{L}}^{2}}$ of the system, with $\varepsilon_{r}$
denoting the reversible COP. This inequality captures the fundamental
cost of operating in finite time and generalizes earlier findings
from low-dissipation thermodynamic models.

Geometric approaches in thermodynamics---such as those based on thermodynamic
length---have proven powerful in quantifying dissipation and optimizing
control protocols \cite{AbiusoEntropy2020,Miller2020,Mehboudi2022,Scandi2018,Scandi2019,Ma2020,Chen2021}.
In this work, we apply these tools to a finite-time quantum tricycle
model and uncover geometric bounds on its performance. By linking
slow-driving perturbation theory with thermodynamic trade-offs, we
derive fundamental constraints on the COP and cooling rate. Our results
bridge microscopic insights from open quantum systems with the macroscopic
thermodynamic principles governing three-heat-source machines, yielding
both fundamental theoretical advances and practical design principles
for efficient quantum thermal devices. This synthesis not only deepens
the understanding of performance limits in quantum refrigerators and
engines but also paves the way for optimized protocols that leverage
quantum effects, crucial for emerging quantum technologies operating
under realistic finite-time constraints.

\begin{figure}
\centering{}\includegraphics[scale=0.25]{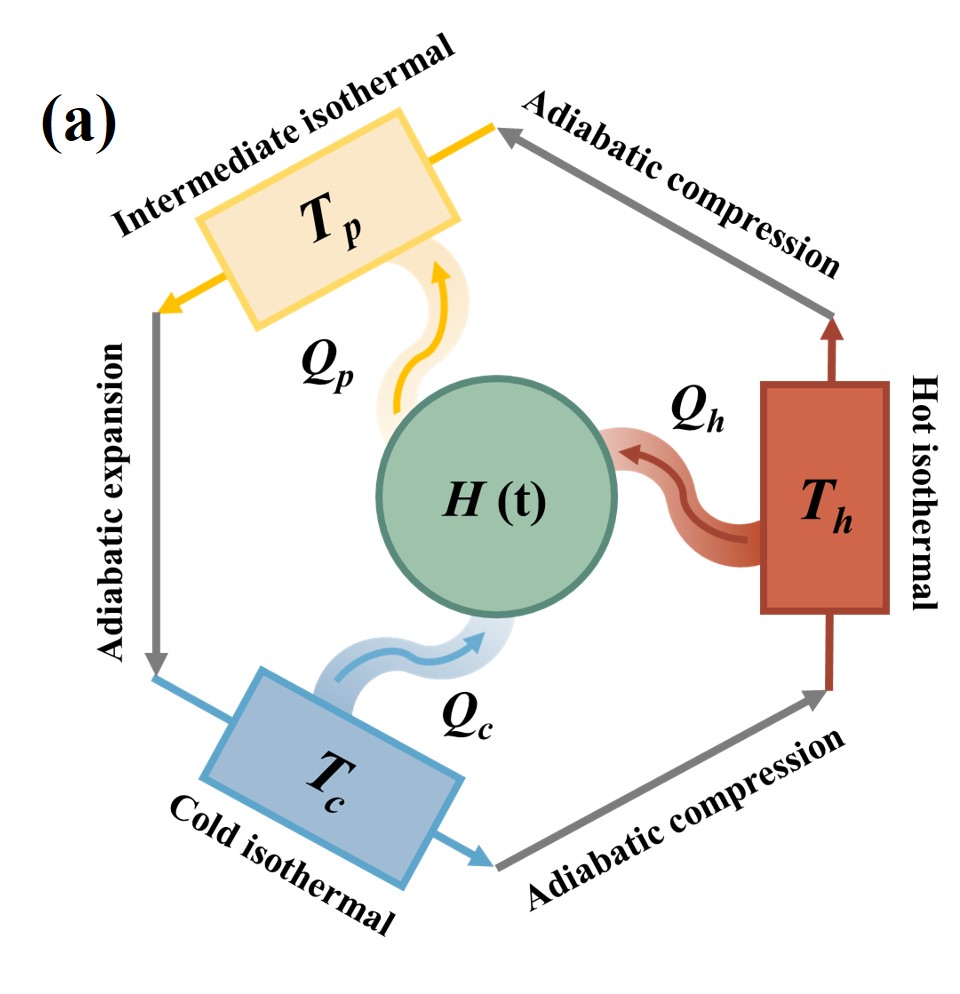}\includegraphics[scale=0.25]{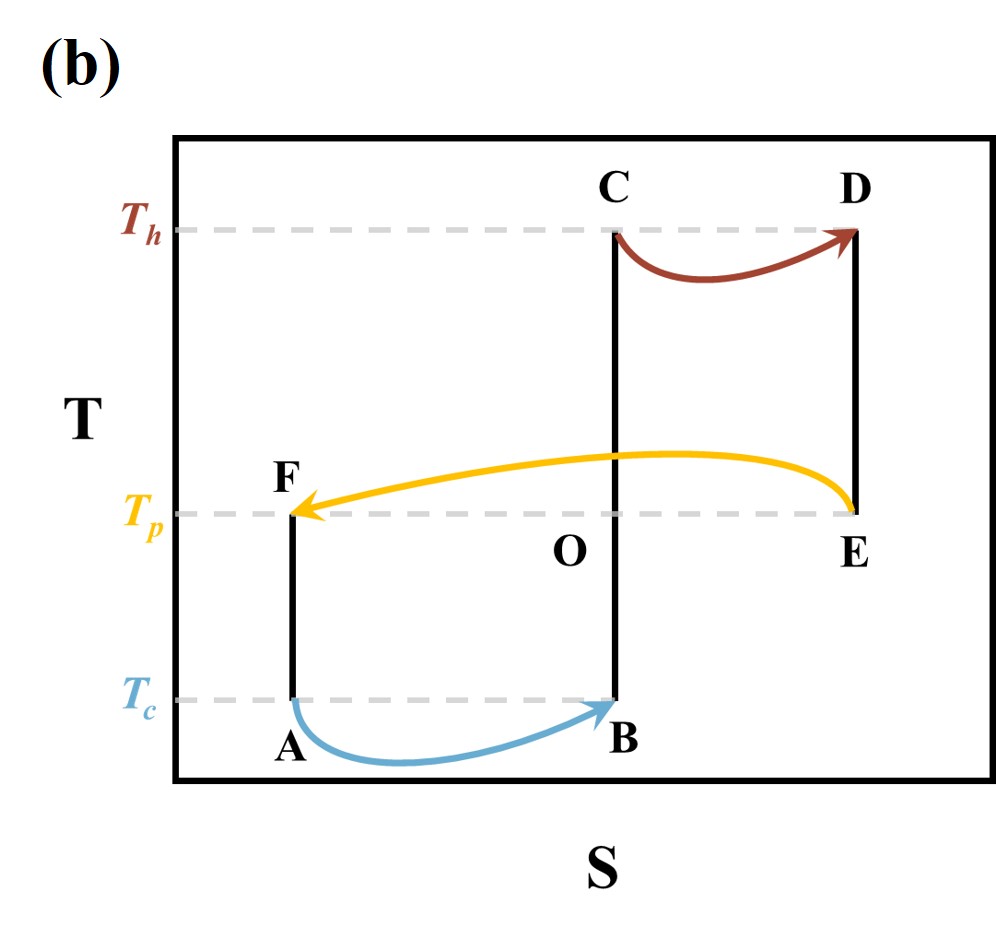}\caption{(a) Schematic representation of a quantum tricycle for refrigeration
and (b)Temperature--entropy ($T\lyxmathsym{–}S$) diagram illustrating
the thermodynamic cycle.}\label{fig:1}
\end{figure}

\textit{Finite-Time Dynamics of the Quantum Tricycle.\label{sec:setup}}---The
tricycle is a fundamental model of a thermal machine operating between
three distinct heat reservoirs, typically labeled as hot $h$, cold
$c$ , and an additional auxiliary (or work) reservoir $p$. Extending
beyond conventional two-reservoir engines, it enables energy exchange
among all three baths, thereby facilitating simultaneous refrigeration,
heat pumping, or power generation. Here, we first propose an extension
of this versatile cycle to the finite-time quantum regime. Similar
to its classical counterpart, the quantum tricycle operates through
a closed cycle in which the working substance sequentially interacts
with the reservoirs $v\in\left\{ h,c,p\right\} $, undergoing three
isothermal processes and three adiabatic processes. Here, we focus
on the quantum refrigeration tricycle, with the model and corresponding
heat flows illustrated in Fig \ref{fig:1}(a). A complete quantum
tricycle \cite{Guo2019} is realized through a six-step sequence:
heat exchange with the cold reservoir $c$ ($A\rightarrow B$), adiabatic
expansion ($B\rightarrow C$), heat exchange with the hot reservoir
$h$ ($C\rightarrow D$), first adiabatic compression ($D\rightarrow E$),
heat exchange with the auxiliary reservoir $p$ ($E\rightarrow F$),
and second adiabatic compression ($F\rightarrow A$). 

During the isothermal process, the Hamiltonian $H(t)$ of the working
substance is time-dependent, and its density operator $\rho(t)$ evolves
according to the Markovian master equation, i.e., $\frac{d}{dt}\rho(t)=\mathcal{L}_{v}(t)[\rho(t)]$,
where $\mathcal{L}_{v}(t)$ denotes the time-dependent quantum Liouvillian
superoperator governing the system's dynamics. By assuming that $dH/dt$
is finite but sufficiently small, and introducing the dimensionless
time-rescaled parameter $s=t/\tau_{v}(s\in[0,1])$ for a given reservoir
$v$, the first-order approximation of the solution of $\rho(t)$
is given by:

\begin{equation}
\tilde{\rho}(s)=\tilde{\rho}_{\mathrm{eq},v}(s)+\frac{1}{\tau_{v}}\tilde{\mathcal{L}}_{v}^{-1}(s)\frac{d}{ds}\left[\tilde{\rho}_{\mathrm{eq},v}(s)\right].\label{eq:rhos}
\end{equation}
Here, $\ensuremath{\tilde{\rho}(s)\equiv\rho\left(\tau_{v}s\right)}$
and $\tilde{\mathcal{L}}_{v}^{-1}(s)\equiv\mathcal{L}_{v}^{-1}\left(\tau_{v}s\right)$
, where $\mathcal{L}_{v}^{-1}\left(\tau_{v}s\right)$ denotes the
Drazin inverse of $\mathcal{L}_{v}(t)$ \cite{Chen2021,Mandal2016}.
The instantaneous Gibbs state $\tilde{\rho}_{\text{eq },v}(s)=\exp\left[-\tilde{H}(s)/\left(k_{B}T_{v}\right)\right]/\mathrm{Tr}\left\{ \exp\left[-\tilde{H}(s)/\left(k_{B}T_{v}\right)\right]\right\} $,
with $\tilde{H}(s)=H\left(\tau_{v}s\right)$. The entropy of the working
substance is given by the von Neumann expression $S(t)=-\mathrm{Tr}\{\rho(t)\ln[\rho(t)]\}=-\mathrm{Tr}\{\tilde{\rho}(s)\ln[\tilde{\rho}(s)]\}$,
which quantifies the degree of mixedness and encodes the system’s
uncertainty or disorder. Based on the definition of heat $Q_{v}=\int_{0}^{\tau}\operatorname{Tr}\left\{ \dot{\rho}(t)H(t)\right\} ]dt$
and Eq. (\ref{eq:rhos}), the amount of heat entering the system from
reservoir $v$ over the interval $\left[0,\tau_{v}\right]$ is given
by

\begin{equation}
Q_{v}=\beta_{v}^{-1}\left(\Delta S_{\mathrm{eq},v}+\Sigma_{v}/\tau_{v}\right)=Q_{v}^{0}+Q_{v}^{1}\label{eq:heat}
\end{equation}
where $\beta_{v}=1/\left(k_{B}T_{v}\right)$, with $k_{B}$ denoting
the Boltzmann constant and $T_{v}$ the temperature of reservoir $v$.
The zeroth order approximation of the density operator $\tilde{\rho}_{\mathrm{eq},v}(s)$
recovers the standard formula for heat in equilibrium thermodynamics,
i.e., $Q_{v}^{0}=\beta_{v}^{-1}\Delta S_{\mathrm{eq},v}$, where
\begin{equation}
\Delta S_{\mathrm{eq},v}=S_{\mathrm{eq},v}\left(\tau_{v}\right)-S_{\mathrm{eq},v}(0)
\end{equation}
is the entropy change of the system's equilibrium state, with
\begin{eqnarray}
S_{\mathrm{eq},v}(t) & = & -\mathrm{Tr}\left\{ \rho_{\mathrm{eq},v}(t)\ln\rho_{\mathrm{eq},v}(t)\right\} \nonumber \\
 & = & -\mathrm{Tr}\left\{ \tilde{\rho}_{\mathrm{eq},v}(s)\ln\tilde{\rho}_{\mathrm{eq},v}(s)\right\} .
\end{eqnarray}
 The first order irreversible corrections of heat $Q_{v}^{1}=\beta_{v}^{-1}\Sigma_{v}/\tau_{\nu}$
with 
\begin{equation}
\Sigma_{v}=\beta_{v}\int_{0}^{1}ds\mathrm{Tr}\left[\tilde{H}(s)\frac{d}{ds}\left\{ \tilde{\mathcal{L}}_{v}^{-1}(s)\frac{d}{ds}\left[\tilde{\rho}_{\mathrm{eq},v}(s)\right]\right\} \right],
\end{equation}
Notably, the first-order irreversible corrections to heat satisfy
$Q_{v}^{1}\leq0$ \cite{AbiusoEntropy2020}. The sum of the first-order
irreversible corrections of heat, represented by the expression $\sum_{v}Q_{v}^{1}$,
is less than or equal to zero. In addition, according to the principles
of energy conversion, the sum of the heat entering a system in a cycle
$\sum_{v}Q_{v}=\sum_{v}Q_{v}^{0}+\sum_{v}Q_{v}^{1}=0$. Thus, the
sum of the zeroth order approximation of heat $\sum_{v}Q_{v}^{0}\geq0$.
In a finite time cycle, it is true that $\sum_{v}Q_{v}^{1}\neq$ 0
, which implies that $\sum_{v}Q_{v}^{0}=\sum_{v}\beta_{v}^{-1}\Delta S_{\mathrm{eq},v}>0$.

\begin{figure*}[!t]
\centering
\begin{centering}
\includegraphics[scale=0.24]{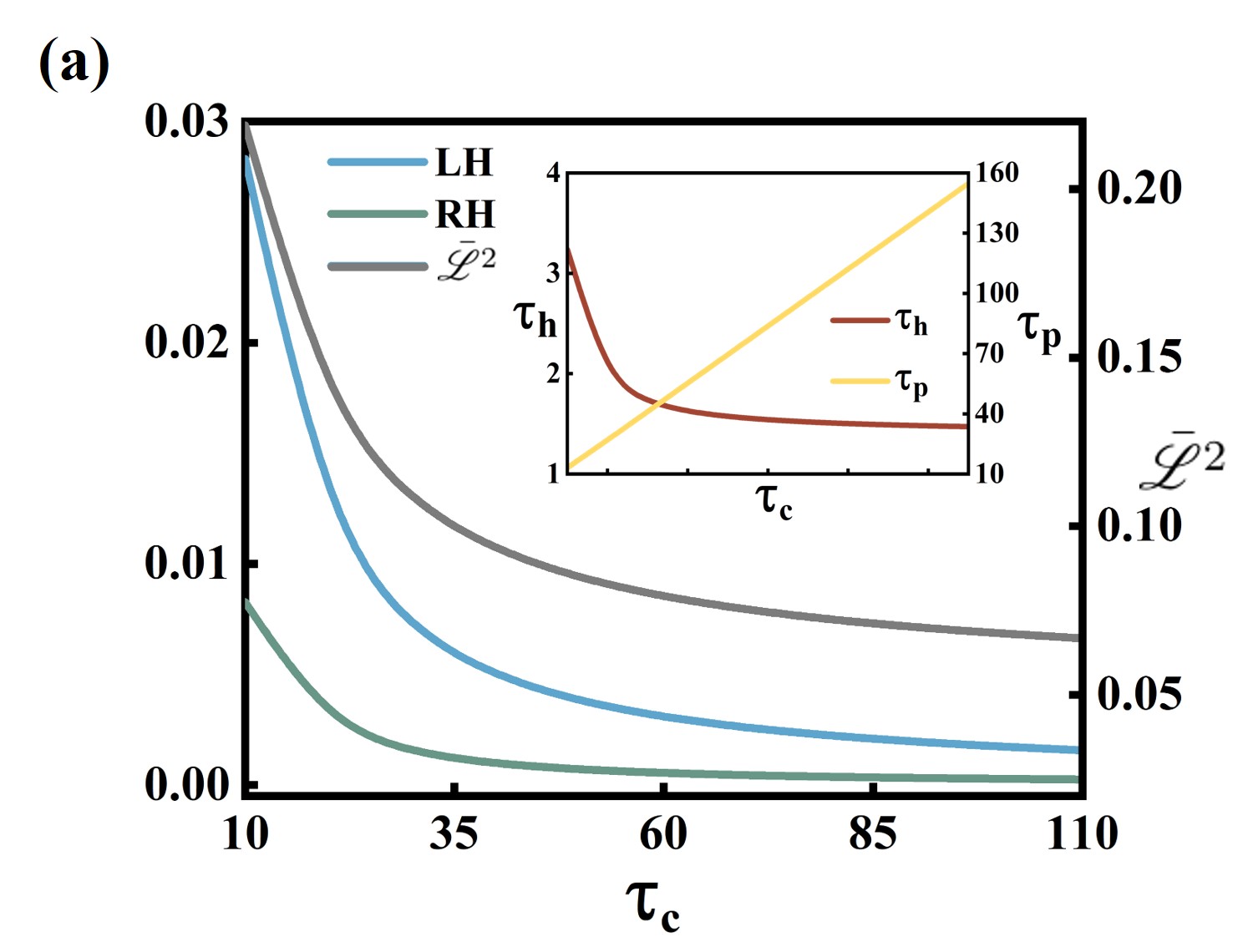}\includegraphics[scale=0.24]{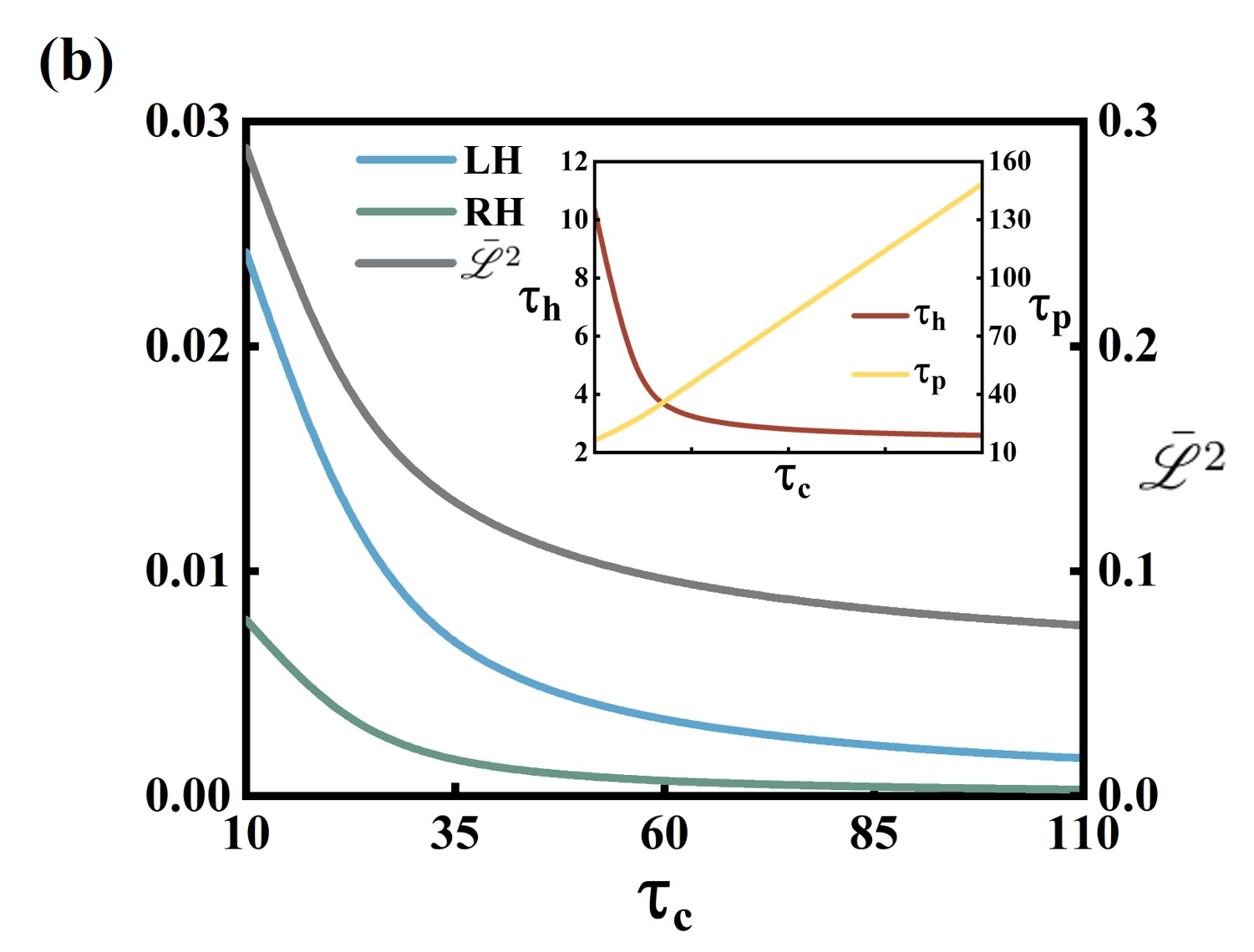}\includegraphics[scale=0.24]{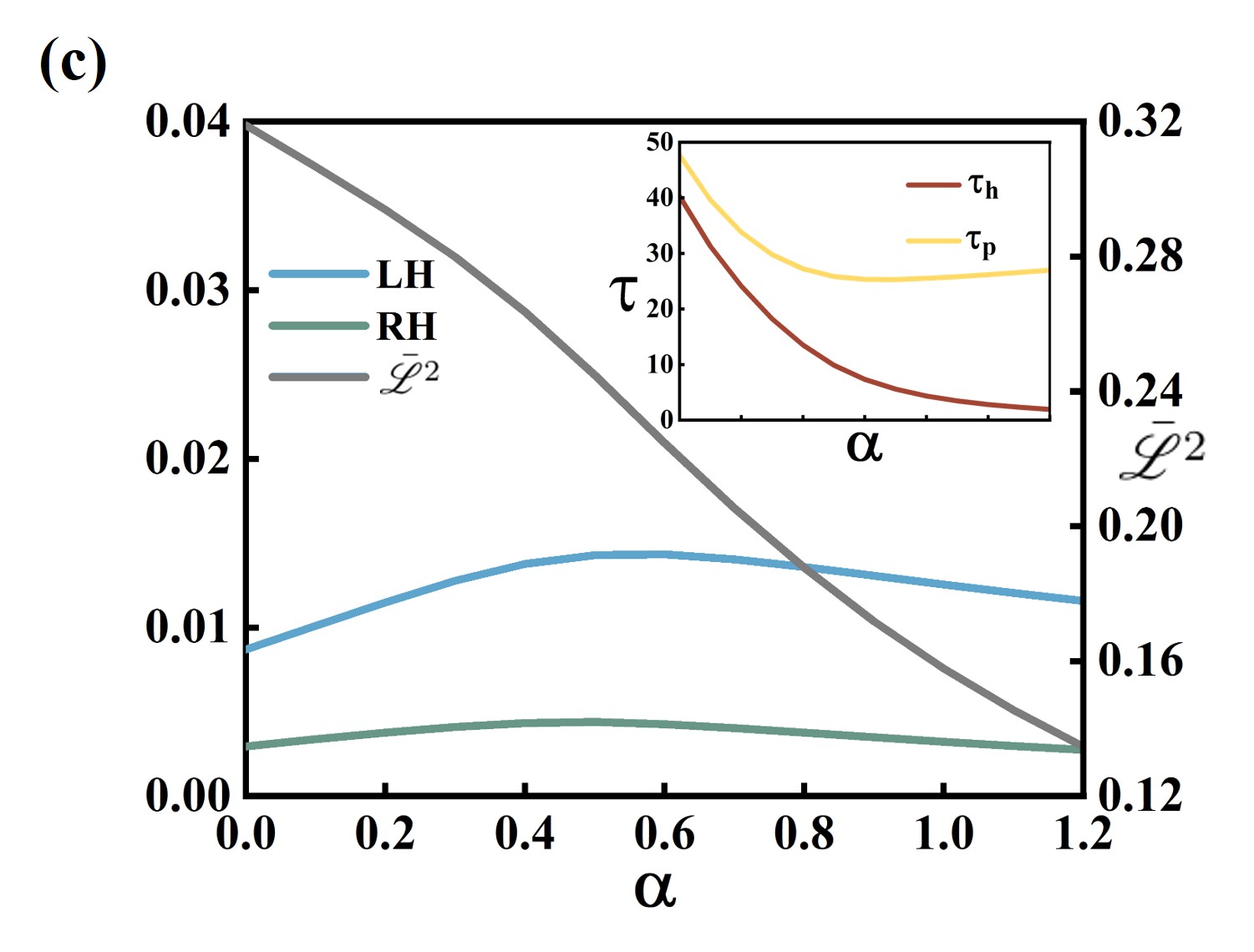}
\par\end{centering}
\caption{ Numerical verification of the two-level system. We plot the reduced
thermodynamic length squared $\bar{\mathcal{\mathscr{L}}^{2}}$, the
left-hand side (LH) defined as $R\left(\varepsilon_{r}/\varepsilon-1\right)$,
and the right-hand side (RH) defined as $\bar{\mathcal{\mathscr{L}}^{2}}/\tau$,
as functions of the contact time $\tau_{c}$ with the cold reservoir.
Figs (a) and (b) correspond to spectral parameters $\alpha=0.8$ (sub-Ohmic)
and $\alpha=1.2$ (super-Ohmic), respectively, with insets showing
the optimized durations $\tau_{h}$ and $\tau_{p}$ as $\tau_{c}$
varies. Fig. (c) displays the dependence of $\bar{\mathcal{\mathscr{L}}^{2}}$,
LH, and RH on $\alpha$ at fixed $\tau_{c}=20$, with the inset illustrating
the optimized values of $\tau_{h}$ and $\tau_{p}$ as $\alpha$ changes.
The remaining parameters are fixed as $\gamma_{0}=k_{B}=\hbar=1$,
$\delta_{c}=k_{B}T_{c}$, $\zeta_{c}=\zeta_{h}=2$, $T_{h}=6$, $T_{p}=2.4$,
and $T_{c}=2$.}\label{fig:2}
\end{figure*}

\textit{Main results.---\label{sec:mainresult}}In this work, we
explore the geometric bounds for trade-off relations in quantum tricycles,
specifically focusing on the implications of thermodynamics in slow-driven
dynamics. By examining the interplay between thermodynamic length
and thermodynamic processes, we derive fundamental limits on the cooling
rate and COP of quantum tricycles. Our findings demonstrate that the
geometric framework serves as a powerful tool for understanding the
trade-offs between the energy dissipation and speed of thermodynamic
processes.

We begin by deriving the equality:

\begin{equation}
R(\frac{\varepsilon_{r}}{\varepsilon}-1)=\frac{\Delta S_{\mathrm{en}}}{(\beta_{c}-\beta_{p})\tau}\label{eq:tradeoff}
\end{equation}
where $\Delta S_{\mathrm{en}}=-\beta_{h}Q_{h}-\beta_{c}Q_{c}-\beta_{p}Q_{p}$
represents the entropy change of the environment. Then by combining
Cauchy-Schwartz inequality, we derived the upper bound for the right
side then we obtain our result as Eq. (\ref{eq:bound}). The detailed
derivation will be shown in \cite{sup}. In the finite-time regime,
the COP $\varepsilon$ can be calculated as 
\begin{equation}
\varepsilon=\frac{Q_{c}}{Q_{h}}=\frac{T_{c}\left[\Delta S_{\mathrm{eq},c}+\Sigma_{c}/\tau_{c}\right]}{T_{h}\left[\Delta S_{\mathrm{eq},h}+\Sigma_{h}/\tau_{h}\right]}.
\end{equation}

In the case of a reversible cycle, a heat exchange process occurs
over an infinitely long duration, i.e., $\tau_{v}\rightarrow\infty$,
leading to $\sum_{v}Q_{v}^{1}$ approaches zero. These quasi-static
isothermal processes ensure that the system remains in thermal equilibrium
with the heat source at all times. Based on Eq. (\ref{eq:heat}) and
the principle of energy conversion, it follows that $\sum_{v}Q_{v}^{0}$
must be equal to zero. In the quasistatic limit, the COP of the reversible
cycle simplifies to

\begin{equation}
\varepsilon_{\mathrm{r}}=\frac{Q_{c}^{0}}{Q_{h}^{0}}=\frac{T_{c}\Delta S_{\mathrm{eq},c}}{T_{h}\Delta S_{\mathrm{eq},h}}=\frac{T_{c}\left(T_{h}-T_{p}\right)}{T_{h}\left(T_{p}-T_{c}\right)}.
\end{equation}

The cooling rate $R$ is defined as the heat extracted $Q_{c}$ from
reservoir $c$ divided by the total time $\tau=\tau_{c}+\tau_{h}+\tau_{p}$
required for cooling, and is expressed as
\begin{equation}
R=\frac{Q_{c}}{\tau}=\frac{\left[\Delta S_{\mathrm{eq},c}+\Sigma_{c}/\tau_{c}\right]}{\beta_{c}\tau}.
\end{equation}

On the right side of Eq. (\ref{eq:bound}), $\bar{\mathcal{\mathscr{L}}^{2}}=\sum_{v}\beta_{v}\mathcal{\mathscr{L}}_{v}^{2}/\tau_{v}\left(\beta_{c}-\beta_{p}\right)$
denotes the reduced thermodynamic length of the cycle. Here,
\begin{equation}
\mathcal{\mathscr{L}}_{v}=\int_{0}^{1}ds\sqrt{\mathrm{Tr}\left\{ \frac{\partial\tilde{H}(s)}{\partial s}\tilde{\mathcal{L}}_{v}^{-1}(s)\frac{d}{ds}\tilde{\rho}_{\mathrm{eq},v}(s)\right\} }
\end{equation}
defines the thermodynamic length \cite{Scandi2019,Chen2021} associated
with the process of the system interacting with reservoir $v$. 

\textit{Numerical simulations.---}The working substance is a two
level system (TLS) with time-dependent Hamiltonian $H(t)=\hbar\omega_{v}(t)\sigma_{z}/2$,
where $\omega_{v}(t)$ is the energy splitting at time $t$, $\sigma_{z}$
is the Pauli matrix in $z$ direction, and $\hbar$ is Planck's constant.
During the isothermal processes in contact with the cold and hot reservoirs
($v=c,h$), the frequency of the TLS is driven by a time-modulated
field according to $\omega_{v}(t)=\delta_{v}\left[\cos\pi\left(t/\tau\right)+\zeta_{v}\right]$
over the time interval $t\in$ $\left[0,\tau\right]$, where $\delta_{v}$
and $\zeta_{v}$ denote the parameters of the amplitude and the displacement,
respectively. For the isothermal process in contact with the intermediate-temperature
reservoir $p$, the system’s frequency is slowly varied according
to $\omega_{p}(t)=$ $\delta_{p}\left[\cos\pi\left(1-t/\tau_{p}\right)+\zeta_{p}\right]$
over the time interval $t\in\left[0,\tau_{p}\right]$. For a given
heat exchange process, the time derivatives of $\omega_{v}(t)$ at
the beginning and the end of time are equal to zero, i.e., $\dot{\omega}_{v}(0)=\dot{\omega}_{v}\left(\tau_{v}\right)=0$.
This also gives rise to $\frac{d}{ds}\left[\tilde{\rho}_{\mathrm{eq},v}(s)\right]=0$
at $s=0$ and $s=1$. According to Eq. (\ref{eq:rhos}), it guarantees
that the system maintains at the equilibrium state both at the beginning
and at the end of the heat exchange process. Furthermore, it ensures
that the system remains close to the instantaneous steady state throughout
the entire cycle. The selection of the scale parameter in frequency
alteration during adiabatic operations enables a smooth transition
of the system between the instantaneous equilibrium states associated
with different reservoirs. With the given temperatures of $T_{c},T_{h}$,
and $T_{p}$, as well as the values of $\zeta_{c},\zeta_{h}$, and
$\delta_{c}$, we can deduce the following relationships \cite{sup}:
$\zeta_{p}=\frac{1+\zeta_{c}\zeta_{h}}{\zeta_{c}+\zeta_{h}},\delta_{h}=\frac{T_{h}\left(\zeta_{c}-1\right)}{T_{c}\left(1+\zeta_{h}\right)}\delta_{c}$,
and $\delta_{p}=\frac{T_{p}\left(\zeta_{c}+\zeta_{h}\right)}{T_{c}\left(1+\zeta_{h}\right)}\delta_{c}$
.

In our regime, the evolution of the TLS during the isothermal process
is governed by the following master equation: 
\begin{eqnarray}
\frac{d}{dt}\rho(t) & = & -\frac{i}{\hbar}[H(t),\rho(t)]\nonumber \\
 & + & \gamma_{v}(t)\left(n_{v}(t)+1\right)\left[\sigma_{-}\rho(t)\sigma_{+}-\frac{1}{2}\left\{ \sigma_{+}\sigma_{-},\rho(t)\right\} \right]\nonumber \\
 & + & \gamma_{v}(t)n_{v}(t)\left[\sigma_{+}\rho(t)\sigma_{-}-\frac{1}{2}\left\{ \sigma_{-}\sigma_{+},\rho(t)\right\} \right]\label{eq:meqtls}
\end{eqnarray}
In this expression, the damping rate $\gamma_{v}(t)=\gamma_{0}\left(\omega_{v}(t)\right)^{\alpha}$
relies on the coupling constant $\gamma_{0}$. The frequency exponent
$\alpha$ is determined by the spectral density $J\left(\omega_{v}(t)\right)\propto\left[\omega_{v}(t)\right]^{\alpha}$
of the bath, which is assumed to be identical for all three bath.
The value of $\alpha$ determines the dissipative nature of the bath,
categorizing it as flat $(\alpha=0)$, sub-Ohmic $(0<\alpha<1)$,
Ohmic $(\alpha=1)$, or super-Ohmic $(\alpha>1)$ \cite{Cavina2017,Cangemi2018}.
The quantity $n_{v}(t)=\left\{ \exp\left[\beta_{v}\hbar\omega_{v}(t)\right]-1\right\} ^{-1}$
represents the mean number of photons associated with bath $v$ at
frequency $\omega_{v}(t)$. The notation $\{A,B\}$ denotes the anticommutator
of the two operators. The operators $\sigma_{+}$ and $\sigma_{-}$correspond
to the raising and lowering operators, respectively.

For a TLS, based on Eq. (\ref{eq:meqtls}), the specific expression
of Liouvillian superoperator $\mathcal{L}_{v}(t)$, instantaneous
equilibrium state $\rho_{\mathrm{eq},v}(t)$, and Drazin inverse $\mathcal{L}_{v}^{-1}(t)$
of the dissipation can be obtained. Thus, the absorbed heat $Q_{v}^{0}$
from bath $v$ can be estimated in the quasi-static limit. Furthermore,
the first-order correction to heat $Q_{v}^{1}$ generated by the irreversible
finite-time process can be calculated as:

\begin{equation}
Q_{v}^{1}=\frac{\hbar}{2\tau_{v}}\int_{0}^{1}\tilde{\omega}_{v}(s)\mathrm{Tr}\left\{ \sigma_{z}\frac{d}{ds}\left\{ \tilde{\mathcal{L}}_{v}^{-1}(s)\frac{d}{ds}\left[\tilde{\rho}_{\mathrm{eq},v}(s)\right]\right\} \right\} ds
\end{equation}

Figure 2 presents our numerical results. We define the left-hand side
$(\mathrm{LH})$ of Eq. (\ref{eq:bound}) as $\mathrm{LH}=R\left(\varepsilon_{r}/\varepsilon-1\right)$,
and the right-hand side $(\mathrm{RH})$ as $\mathrm{RH}=L^{2}/\tau$.
Both quantities are plotted as functions of the contact time $\tau_{c}$
with the low temperature reservoir and the spectral parameter $\alpha$.
The durations $\tau_{h}$ and $\tau_{p}$ are optimized using the
Lagrange multiplier method \cite{sup}.

Figures 2(a) and 2(b) illustrate the cases of $\alpha=0.8$ and $\alpha=1.2$,
corresponding to the sub-Ohmic and super-Ohmic regimes, respectively.
In both cases, the value of $\bar{\mathcal{\mathscr{L}}^{2}}$ decreases
as $\tau_{c}$ increases. Notably, as $\tau_{c}$ grows, the thermodynamic
bound becomes tighter, with $\tau_{p}$ showing a monotonically increasing
trend and $\tau_{h}$ decreasing rapidly before stabilizing. These
behaviors, combined with our optimization strategy, suggest that the
system achieves its maximum cooling rate at a given coefficient of
performance (COP) when it maintains prolonged contact with the low-
and medium temperature reservoirs, while minimizing interaction time
with the high-temperature reservoir.

Furthermore, the bound tightens for a fixed value of $\alpha$. As
shown in Fig. 2(c), the difference between LH and RH initially increases
and then decreases as $\alpha$ grows. The optimized value of $\tau_{p}$
exhibits a non-monotonic behavior-first decreasing and then increasing
with $\alpha$ while $\tau_{h}$ gradually declines. Additionally,
the variation of $\bar{\mathcal{\mathscr{L}}^{2}}$ with respect to
$\alpha$ reveals a consistent decreasing trend, indicating reduced
thermodynamic length in more dissipative environments.

\textit{Conclusion.---}In summary, we introduced a finite-time quantum
tricycle model and analyzed its performance under slow driving perturbations.
By expanding the heat exchanged during isothermal processes with respect
to time, we identified fundamental geometric constraints that govern
the trade-offs between energy dissipation and process speed. Our results
demonstrate that the geometric framework provides a powerful tool
for characterizing performance limits in quantum thermal machines.
These findings deepen the understanding of quantum thermodynamics
and offer guiding principles for the design of efficient quantum devices.\\

\textit{Acknowledgement.---}This work has been supported by the Natural
Science Foundation of Fujian Province (2023J01006), National Natural
Science Foundation of China (12364008 and 12365006), and Fundamental
Research Fund for the Central Universities (20720240145).\\


\begin{thebibliography}{10}
\bibitem{Broeck2005}C. Van den Broeck, Thermodynamic efficiency at
maximum power, Phys. Rev. Lett. 95, 190602 (2005).

\bibitem{Curzon1975}F. L. Curzon and B. Ahlborn, Efficiency of a
Carnot engine at maximum power output, Am. J. Phys. 43, 22 (1975).

\bibitem{Salamon1980}P. Salamon, A. Nitzan, B. Andresen, and R. S.
Berry, Minimum entropy production and the optimization of heat engines,
Phys. Rev. A 21, 2115 (1980).

\bibitem{Sekimoto1997}K. Sekimoto and S. Sasa, Complementarity relation
for irreversible process derived from stochastic energetics, J. Phys.
Soc. Jpn. 66, 3326 (1997).

\bibitem{Esposito2009}M. Esposito, K. Lindenberg, and C. Van den
Broeck, Thermoelectric efficiency at maximum power in a quantum dot,
Europhys. Lett. 85, 60010 (2009).

\bibitem{Tu2008}Z. Tu, Efficiency at maximum power of Feynman's ratchet
as a heat engine, J. Phys. A: Math. Theor. 41, 312003 (2008).

\bibitem{Izumida2009}Y. Izumida and K. Okuda, Onsager coefficients
of a finite-time Carnot cycle, Phys. Rev. E 80, 021121 (2009).

\bibitem{Izumida2010}Y. Izumida and K. Okuda, Onsager coefficients
of a Brownian Carnot cycle, Eur. Phys. J. B 77, 499 (2010).

\bibitem{Miller2019}H. J. D. Miller, M. Scandi, J. Anders, and M.
Perarnau-Llobet, Work fluctuations in slow processes: quantum signatures
and optimal control, Phys. Rev. Lett. 123, 230603 (2019).

\bibitem{Kosloff2013}R. Kosloff, Quantum thermodynamics: A dynamical
viewpoint, Entropy 15, 2100 (2013).

\bibitem{Guo2019}J. Guo, H. Yang, H. Zhang, J. Gonzalez-Ayala, J.
Roco, A. Medina, and A. C. Hernández, Thermally driven refrigerators:
Equivalent low-dissipation three-heat-source model and comparison
with experimental and simulated results, Energy Convers. Manage. 198,
111917 (2019).

\bibitem{Yan1989}Z. Yan and J. Chen, An optimal endoreversible three-heat-source
refrigerator, J. Appl. Phys. 65, 1 (1989).

\bibitem{Chen1989}J. Chen and Z. Yan, Equivalent combined systems
of three-heat-source heat pumps, J. Chem. Phys. 90, 4951 (1989).

\bibitem{Chen1989PRA}J. Chen and Z. Yan, Unified description of endoreversible
cycles, Phys. Rev. A 39, 4140 (1989).

\bibitem{Schmiedl2007}T. Schmiedl and U. Seifert, Efficiency at maximum
power: An analytically solvable model for stochastic heat engines,
Europhys. Lett. 81, 20003 (2007).

\bibitem{Abiuso2020}P. Abiuso and M. Perarnau-Llobet, Optimal cycles
for low-dissipation heat engines, Phys. Rev. Lett. 124, 110606 (2020).

\bibitem{Scandi2018}M. Scandi, Quantifying dissipation via thermodynamic
length, Ph.D. thesis, Ludwig-Maximilians-Universität München (2018).

\bibitem{Esposito2010}M. Esposito, R. Kawai, K. Lindenberg, and C.
Van den Broeck, Quantum-dot Carnot engine at maximum power, Phys.
Rev. E 81, 041106 (2010).

\bibitem{Tomas2012}C. De Tomás, A. C. Hernández, and J. M. M. Roco,
Optimal low symmetric dissipation Carnot engines and refrigerators,
Phys. Rev. E 85, 010104 (2012).

\bibitem{Wang2012}Y. Wang, M. Li, Z. Tu, A. C. Hernández, and J.
Roco, Coefficient of performance at maximum figure of merit and its
bounds for low-dissipation Carnot-like refrigerators, Phys. Rev. E
86, 011127 (2012).

\bibitem{Gonzalez-Ayala2018}J. Gonzalez-Ayala, A. Medina, J. Roco,
and A. C. Hernández, Entropy generation and unified optimization of
Carnot-like and low-dissipation refrigerators, Phys. Rev. E 97, 022139
(2018).

\bibitem{Holubec2020}V. Holubec and Z. Ye, Maximum efficiency of
low-dissipation refrigerators at arbitrary cooling power, Phys. Rev.
E 101, 052124 (2020).

\bibitem{Ye2021}Z. Ye and V. Holubec, Maximum efficiency of absorption
refrigerators at arbitrary cooling power, Phys. Rev. E 103, 052125
(2021).

\bibitem{Tomas2013}C. De Tomás, J. Roco, A. C. Hernández, Y. Wang,
and Z. Tu, Low-dissipation heat devices: Unified trade-off optimization
and bounds, Phys. Rev. E 87, 012105 (2013).

\bibitem{Miller2021}H. J. D. Miller, M. H. Mohammady, M. Perarnau-Llobet,
and G. Guarnieri, Thermodynamic uncertainty relation in slowly driven
quantum heat engines, Phys. Rev. Lett. 126, 210603 (2021).

\bibitem{Pietzonka2018}P. Pietzonka and U. Seifert, Universal trade-off
between power, efficiency, and constancy in steady-state heat engines,
Phys. Rev. Lett. 120, 190602 (2018).

\bibitem{key-76}Y. Hasegawa, Quantum thermodynamic uncertainty relation
for continuous measurement, Phys. Rev. Lett. 125, 050601 (2020).

\bibitem{key-79}T. Van Vu and K. Saito, Thermodynamics of precision
in Markovian open quantum dynamics, Phys. Rev. Lett. 128, 140602 (2022).

\bibitem{key-57}A.C. Barato and U. Seifert, Thermodynamic uncertainty
relation for biomolecular processes, Phys. Rev. Lett. 114, 158101
(2015).

\bibitem{key-60}J. M. Horowitz and T.R. Gingrich, Thermodynamic uncertainty
relations constrain non-equilibrium fluctuations, Nat. Phys. 16, 15
(2020).

\bibitem{key-66}K. Macieszczak, K. Brandner, and J.P. Garrahan, Unified
thermodynamic uncertainty relations in linear response, Phys. Rev.
Lett. 121, 130601 (2018)

\bibitem{key-70}K. Liu, Z. Gong, and M. Ueda, Thermodynamic uncertainty
relation for arbitrary initial states, Phys. Rev. Lett. 125, 140602
(2020).

\bibitem{key-73}T. Koyuk and U. Seifert, Thermodynamic uncertainty
relation for time-dependent driving, Phys. Rev. Lett. 125, 260604
(2020).

\bibitem{key-80}S.-L. Liang, Y.-H. Ma, D. M. Busiello, and P. D.
L. Rios, Minimal model for Carnot efficiency at maximum power, Phys.
Rev. Lett. 134, 027101 (2025).

\bibitem{Cavina2017}V. Cavina, A. Mari, and V. Giovannetti, Slow
dynamics and thermodynamics of open quantum systems, Phys. Rev. Lett.
119, 050601 (2017).

\bibitem{Breuer2002}H.-P. Breuer and F. Petruccione, The theory of
open quantum systems (Oxford University Press, 2002).

\bibitem{Rivas2012}A. Rivas and S. F. Huelga, Open quantum systems,
Vol. 10 (Springer, 2012).

\bibitem{AbiusoEntropy2020}P. Abiuso, H. J. Miller, M. Perarnau-Llobet,
and M. Scandi, Geometric optimisation of quantum thermodynamic processes,
Entropy 22, 1076 (2020).

\bibitem{Miller2020}H. J. D. Miller and M. Mehboudi, Geometry of
work fluctuations versus efficiency in microscopic thermal machines,
Phys. Rev. Lett. 125, 260602 (2020).

\bibitem{Mehboudi2022}M. Mehboudi and H. J. D. Miller, Thermodynamic
length and work optimization for Gaussian quantum states, Phys. Rev.
A 105, 062434 (2022).

\bibitem{Scandi2019}M. Scandi and M. Perarnau-Llobet, Thermodynamic
length in open quantum systems, Quantum 3, 197 (2019).

\bibitem{Ma2020}Y.-H. Ma, R.-X. Zhai, J. Chen, C. Sun, and H. Dong,
Experimental test of the 1/$\tau$-scaling entropy generation
in finite-time thermodynamics, Phys. Rev. Lett. 125, 210601 (2020).

\bibitem{Chen2021}J.-F. Chen, C. P. Sun, and H. Dong, Extrapolating
the thermodynamic length with finite-time measurements, Phys. Rev.
E 104, 034117 (2021).

\bibitem{Mandal2016}D. Mandal and C. Jarzynski, Analysis of slow
transitions between nonequilibrium steady states, J. Stat. Mech. 2016,
063204 (2016).

\bibitem{Cangemi2018}L. M. Cangemi, G. Passarelli, V. Cataudella,
P. Lucignano, and G. De Filippis, Beyond the Born-Markov approximation:
Dissipative dynamics of a single qubit, Phys. Rev. B 98, 184306 (2018).

\bibitem{sup}Supplementary Materials

\end{thebibliography}
\end{document}


\title{Supplemental Material for ``Geometry Bound for Trade-off Relation
in Quantum Tricycle'' }
\author{Shihao Xia}
\thanks{These authors contributed equally to this work.}
\author{Jingyi Chen}
\thanks{These authors contributed equally to this work.}
\author{Jincan Chen}
\author{Shanhe Su}
\email{sushanhe@xmu.edu.cn}

\affiliation{Department of Physics, Xiamen University, Xiamen, 361005, People's
Republic of China }
\date{\today}

\maketitle
\tableofcontents
\renewcommand{\theequation}{S\arabic{equation}}
\renewcommand{\thetable}{S\arabic{table}}
\renewcommand{\thefigure}{S\arabic{figure}}
\renewcommand{\thesection}{S\arabic{section}}
\bibliographystyle{apsrev4-1}

\section{THE CONTROL PROTOCOL OF A FINITE-TIME QUANTUM TRICYCLE}\label{sec:THE-CONTROL-PROTOCOL}

\subsubsection{Quantum Tricycle Model}

The detail of the control protocols of the quantum refrigeration tricycle
are designed as follows:

(1) Heat exchange with cold reservoir $c$ (from point $A$ to $B$):
When the system is coupled to reservoir $c$ with low temperature
$T_{c}$, the time-modulated field drives the system in the time interval
$t\in$ $\left[0,\tau_{c}\right]$. 

(2) Adiabatic expansion (from point $B$ to $C$): The system is isolated
from any reservoir and the frequency suddenly shifted from $H_{c}\left(\tau_{c}\right)$
to $H_{h}(0)=\left(T_{h}/T_{c}\right)H_{c}\left(\tau_{c}\right)$.
The scaling factor $T_{h}/T_{c}$ is introduced, ensuring that the
system continuously remains in equilibrium state during a reversible
quantum tricycle. In addition, the rapidly changing condition of adiabatic
process prevents the system from adapting its configuration during
the process, and thus the probability density remains unchanged.

(3) Heat exchange with hot reservoir $h$ (from point $C$ to $\mathrm{D})$
: The system is coupled to reservoir $h$ with high temperature $T_{h}$
in the time interval $t\in\left[0,\tau_{h}\right]$. 

(4) First adiabatic compression (from point $D$ to $E$): The system
is decoupled from the reservoir and a sudden quench is performed,
in which the frequency is changed from $H_{h}\left(\tau_{h}\right)$
into $H_{p}(0)=\left(T_{p}/T_{h}\right)H_{h}\left(\tau_{h}\right)$.

(5) Heat exchange with reservoir $p$ (from point $E$ to $F$): An
irreversible heat exchange process follows by coupling the system
with reservoir $p$ at intermediate temperature $T_{p}$. 

(6) Second adiabatic compression (from point $F$ to $A$): Finally,
the system is disconnected from the reservoir and a sudden quench
restores the frequency back to $H_{c}(0)=\left(T_{c}/T_{p}\right)H_{p}\left(\tau_{p}\right)$.

\begin{figure}
\centering{}\includegraphics[scale=0.3]{fig1a}\includegraphics[scale=0.3]{fig1b}\includegraphics[scale=0.3]{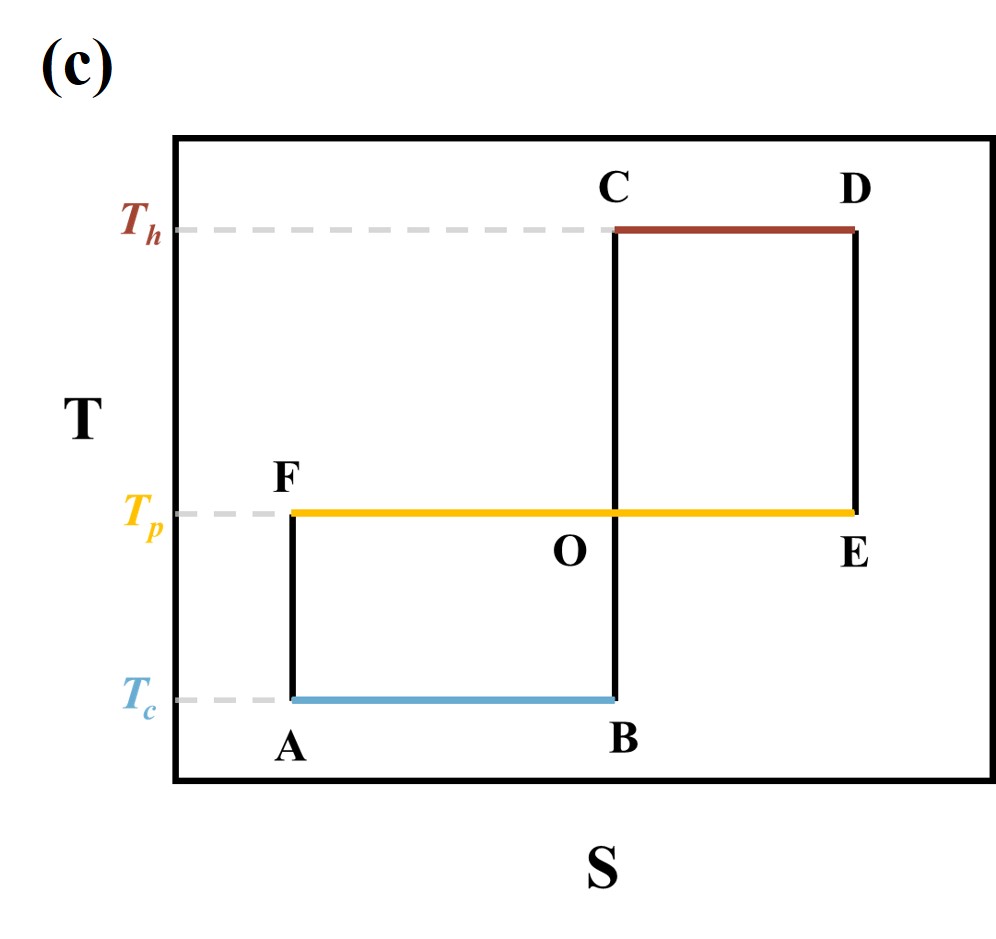}\caption{(a) Schematic representation of a quantum tricycle. (b) The temperature-entropy
diagram of a FTQTC. (c) The temperature-entropy diagram of a reversible
quantum tricycle. }\label{fig:1}
\end{figure}

The density operator $\rho(t)$ of the system evolves according to
the Markovian master equation,

\begin{equation}
\frac{d}{dt}\rho(t)=\mathcal{L}_{v}(t)[\rho(t)].\label{eq:mmeq}
\end{equation}

By considering that the Liouvillian operator $\mathcal{L}_{v}(t)$
obeys the quantum detailed balance with respect to $H(t)$ at all
times, the Gibbs state 
\begin{equation}
\rho_{\mathrm{eq},v}(t)=\exp\left[-\beta_{v}H_{v}(t)\right]/\mathrm{Tr}\left\{ \exp\left[-\beta_{v}H_{v}(t)\right]\right\} \label{eq:Gibbs}
\end{equation}
becomes the immediate stationary state of Eq. (\ref{eq:mmeq}), meaning
that

\begin{equation}
\mathcal{L}_{v}(t)\left[\rho_{\mathrm{eq},v}(t)\right]=0.\label{eq:sseq}
\end{equation}

As the modulation of Hamiltonian becomes infinitely slow, the dynamics
of the system follow a quasi-static isothermal trajectory. During
this trajectory, the system gradually adjusts to the instantaneous
equilibrium state near the Gibbs ensemble $\rho_{\text{eq },v}(t)$
\cite{Cavina2017}. Substituting Eq. (\ref{eq:sseq}) into Eq. (\ref{eq:mmeq}),
we can derive an alternative evolution equation for $\rho(t)$ as
follows:

\begin{equation}
\frac{d}{dt}\rho(t)=\mathcal{L}_{v}(t)\left[\rho(t)-\rho_{\mathrm{eq},v}(t)\right]\label{eq:eveq}
\end{equation}

As $\rho(t)-\rho_{\mathrm{eq},v}(t)$ is a trace-less Herminia operator,
and the action of a super-operator on a traceless subspace is invertible,
the solution of Eq. (\ref{eq:eveq}) can be expressed as follows:

\begin{equation}
\rho(t)=\rho_{\mathrm{eq},v}(t)+\mathcal{L}_{v}^{-1}(t)\left[\frac{d}{dt}\rho_{\mathrm{eq},v}(t)\right],\label{eq:1stexpandrho}
\end{equation}
where $\mathcal{L}_{v}^{-1}(t)$ is the Drazin inverse of $\mathcal{L}_{v}(t)$.
Therefore, the solution for Eq. (\ref{eq:eveq}) can be expanded in
a series form through an iterative process. The expansion is given
by the following expression:

\begin{equation}
\rho(t)\approx\sum_{n=0}^{\infty}\left(\mathcal{L}_{v}^{-1}(t)\frac{d}{dt}\right)^{n}\left[\rho_{\mathrm{eq},v}(t)\right].\label{eq:rhoseries}
\end{equation}

Note that in Eq. (\ref{eq:rhoseries}), we have omitted the term $\left(\mathcal{L}_{v}^{-1}(t)\frac{d}{dt}\right)^{n+1}[\rho(t)]$
with $n$ approaching infinity. This omission is justified by assuming
that the derivative of the density operator with respect to time,
when taken an infinite number of times, becomes zero.

We can rewrite Eq. (\ref{eq:rhoseries}) by introducing the dimensionless
time-rescaled parameter $s=t/\tau_{v}$ as follows

\begin{equation}
\tilde{\rho}(s)=\sum_{n=0}^{\infty}\left(\tilde{\mathcal{L}}_{v}^{-1}(s)\frac{1}{\tau_{v}}\frac{d}{ds}\right)^{n}\left[\tilde{\rho}_{\mathrm{eq},v}(s)\right],\label{eq:tsrho}
\end{equation}
where $\tilde{\rho}(s)\equiv\rho(\tau s)$ and $\tilde{\mathcal{L}}_{v}^{-1}(s)\equiv\mathcal{L}_{v}^{-1}(\tau s)$.
Rescaling the time provides the advantage of incorporating the duration
$\tau_{v}$ of the evolution as a straightforward multiplicative factor
in each term of the sum in Eq. (\ref{eq:tsrho}). The state of the
system now relies on the time $s$ within the unit interval $s\in[0,1]$,
while the impact of the control protocol on the dynamics is encompassed
within $\tilde{\mathcal{L}}_{v}^{-1}(s)$. Hence, Eq. (\ref{eq:tsrho})
provides a perturbation expansion of the solution of the density operator
in terms of powers of $1/\tau_{v}$. During the finite-time slow driving
process, there exists a lag between states $\tilde{\rho}(s)$ and
$\tilde{\rho}_{\text{eq },v}(s)$. Considering the first-order perturbation,
we obtain:

\begin{equation}
\tilde{\rho}(s)=\tilde{\rho}_{\mathrm{eq},v}(s)+\frac{1}{\tau_{v}}\tilde{\mathcal{L}}_{v}^{-1}(s)\frac{d}{ds}\left[\tilde{\rho}_{\mathrm{eq},v}(s)\right].\label{eq:rho1st}
\end{equation}
which is the Eq.(1) in the main text.

\subsubsection{Calculation for Key Quantities}

In the main text, we give the amount of heat entering the system from
reservoir $v$ during the interval $\left[0,\tau_{v}\right]$ is
\begin{equation}
Q_{v}=Q_{v}^{0}+Q_{v}^{1}\label{eq:Qre}
\end{equation}
Next, we provide the specific form for $Q_{v}$. Building upon the
previous derivation, our objective is to investigate how deviations
from the corresponding quasistatic trajectory, as described in Eq.(\ref{eq:rho1st}),
influence the thermodynamic properties of the process. To achieve
this objective, it is important to note that the mean energy and von
Neumann entropy of system can be expressed as $U(t)=\mathrm{Tr}[H(t)\rho(t)]$
and $S(t)=-\mathrm{Tr}[\rho(t)\ln\rho(t)]$, respectively. In a finite-time
process with a duration of $\tau_{v}$, the change in the internal
energy is typically divided into two components\textbf{ }$U\left(\tau_{v}\right)-U(0)=\int_{0}^{\tau_{v}}dt\mathrm{Tr}\left[H(t)\frac{d}{dt}\rho(t)\right]+\int_{0}^{\tau_{v}}dt\mathrm{Tr}\left[\rho(t)\frac{d}{dt}H(t)\right]$.
By employing the definition of heat and considering the first-order
perturbation, the amount of heat transferred into the system from
reservoir $\nu$ during the interval $\left[0,\tau_{v}\right]$ is
given by:

\begin{equation}
Q_{v}=\int_{0}^{\tau_{v}}dt\mathrm{Tr}\left[H_{v}(t)\frac{d}{dt}\rho_{v}(t)\right]=\int_{0}^{1}ds\mathrm{Tr}\left[\tilde{H_{v}}(s)\frac{d}{ds}\tilde{\rho_{v}}(s)\right],\label{eq:Q}
\end{equation}
where the time-rescaled Hamiltonian $\tilde{H}(s)\equiv H\left(\tau_{v}s\right)$.
In a quasi-static process that maintains the equilibrium state between
the system and the bath, the zeroth-order approximation in equilibrium
thermodynamics suggests that the system absorbs an amount of heat
given by the expression: 
\begin{equation}
Q_{v}^{0}=\int_{0}^{\tau_{v}}dt\mathrm{Tr}\left[H_{v}(t)\frac{d}{dt}\rho_{\mathrm{eq},v}(t)\right]=\int_{0}^{1}ds\mathrm{Tr}\left[\tilde{H_{v}}(s)\frac{d}{ds}\left[\tilde{\rho}_{\mathrm{eq},v}(s)\right]\right].\label{eq:Q0}
\end{equation}

Note that the entropy of the system in the equilibrium state is given
by
\begin{equation}
S_{\mathrm{eq},v}(t)=-\mathrm{Tr}\left[\rho_{\mathrm{eq},v}(t)\ln\left[\rho_{\mathrm{eq},v}(t)\right]\right]=-\mathrm{Tr}\left[\tilde{\rho}_{\mathrm{eq},v}(s)\ln\left[\tilde{\rho}_{\mathrm{eq},v}(s)\right]\right].\label{eq:Seq}
\end{equation}

Combining the definition in Eq. (\ref{eq:Gibbs}), we have the relation
between Eqs. (\ref{eq:Q0}) and (\ref{eq:Seq}) 
\begin{equation}
Q_{v}^{0}=\beta_{v}^{-1}\Delta S_{\mathrm{eq},v}\label{eq:Q0S}
\end{equation}
which is associated with the change in entropy $\Delta S_{\mathrm{eq},v}=$
$S_{\mathrm{eq},v}\left(\tau_{v}\right)-S_{\mathrm{eq},v}(0)$ along
the quasi-static trajectory during the time interval $\left[0,\tau_{v}\right]$.
The first-order irreversible correction of heat
\begin{align}
Q_{v}^{1} & =\int_{0}^{\tau_{v}}dt\mathrm{Tr}\left[H_{v}(t)\frac{d}{dt}\left\{ \mathcal{L}_{v}^{-1}(t)\frac{d}{dt}\left[\rho_{\mathrm{eq},v}(t)\right]\right\} \right]\nonumber \\
 & =\tau_{v}^{-1}\int_{0}^{1}ds\mathrm{Tr}\left[\tilde{H}_{v}(s)\frac{d}{ds}\left\{ \tilde{\mathcal{L}}_{v}^{-1}(s)\frac{d}{ds}\left[\tilde{\rho}_{\mathrm{eq},v}(s)\right]\right\} \right]\label{eq:Q1}\\
 & =\beta_{v}^{-1}\Sigma_{v}/\tau_{v}\nonumber 
\end{align}
where
\begin{equation}
\Sigma_{v}=\beta_{v}\int_{0}^{1}ds\mathrm{Tr}\left[\tilde{H}_{v}(s)\frac{d}{ds}\left\{ \tilde{\mathcal{L}}_{v}^{-1}(s)\frac{d}{ds}\left[\tilde{\rho}_{\mathrm{eq},v}(s)\right]\right\} \right]\label{eq:s1}
\end{equation}
expresses the increase in dissipation that deviates from the reversible
limit. Combining Eqs. (\ref{eq:Qre}), (\ref{eq:Q0S}) and (\ref{eq:Q1}),
we can also represent $Q_{v}$ as

\begin{equation}
Q_{v}=\beta_{v}^{-1}\left(\Delta S_{\mathrm{eq},v}+\Sigma_{v}/\tau_{v}\right).\label{eq:Qfin}
\end{equation}

Based on the the temperature-entropy diagram shown in Figs. \ref{fig:1}(b)
and \ref{fig:1}(c),\textcolor{red}{{} }it is observed that $Q_{h}^{0}>0,Q_{c}^{0}>0$,
and $Q_{p}^{0}<0$, as $\Delta S_{\text{eq },h}>0,\Delta S_{\text{eq },c}>0$,
and $\Delta S_{\text{eq},p}<$ 0. In Ref. \cite{AbiusoEntropy2020},
it is demonstrated that the first-order irreversible corrections of
heat, i.e., $Q_{v}^{1}\leq0$. The sum of the first-order irreversible
corrections of heat, represented by the expression $\sum_{v}Q_{v}^{1}$,
is less than or equal to zero. In addition, according to the principles
of energy conversion, the sum of the heat entering a system in a cycle

\begin{equation}
\sum_{v}Q_{v}=\sum_{v}Q_{v}^{0}+\sum_{v}Q_{v}^{1}=0.\label{eq:conversion}
\end{equation}

Thus, the sum of the zeroth order approximation of heat $\sum_{v}Q_{v}^{0}\geq0$.
In a finite time cycle, it is true that $\sum_{v}Q_{v}^{1}\neq$ 0
, which implies that $\sum_{v}Q_{v}^{0}=\sum_{v}\beta_{v}^{-1}\Delta S_{\text{eq},v}>0$.
As a result, the area of rectangle ABOF, i.e., $\left(T_{p}-T_{c}\right)\Delta S_{\text{eq},c}$
is not equal to the area of rectangle CDEO, i.e., $\left(T_{h}-T_{p}\right)\Delta S_{\mathrm{eq},h}$,
as shown in Fig. 1(b).

In the finite-time regime, utilizing the Eq.(\ref{eq:Qfin}), the
COP $\varepsilon$ of the quantum tricycle can be calculated as

\begin{equation}
\varepsilon=\frac{Q_{c}}{Q_{h}}=\frac{T_{c}\left[\Delta S_{\mathrm{eq},c}+\Sigma_{c}/\tau_{c}\right]}{T_{h}\left[\Delta S_{\mathrm{eq},h}+\Sigma_{h}/\tau_{h}\right]},\label{eq:cop}
\end{equation}
while the cooling rate is determined by the heat $Q_{c}$ entering
the system from reservoir $c$ divided by the total time required
for cooling:

\begin{equation}
R=\frac{Q_{c}}{\tau}=\frac{\left[\Delta S_{\mathrm{eq},c}+\Sigma_{c}/\tau_{c}\right]}{\beta_{c}\tau}.\label{eq:cr}
\end{equation}
Here, $\tau=\tau_{c}+\tau_{h}+\tau_{p}$, and the duration of the
adiabatic process is significantly shorter than that of the heat exchange
process and can be considered negligible. 

In the case of a reversible cycle, a heat exchange process occurs
over an infinitely extended duration, i.e., $\tau_{v}\rightarrow\infty$,
leading to $\sum_{v}Q_{v}^{1}$ approaches zero. These quasi-static
isothermal processes ensure that the system remains in thermal equilibrium
with the reservoir at all times. It follows from Eq. (\ref{eq:conversion})
that $\sum_{v}Q_{v}^{0}$ must be equal to zero \cite{Chen1989}.
This leads to the equality between the area of rectangle ABOF and
that of rectangle CDEO, as depicted in Fig. 1(c). At the quasistatic
limit, and the coefficient of performance (COP) of the reversible
cycle is then simplified as follows

\begin{equation}
\varepsilon_{\mathrm{r}}=\frac{Q_{c}}{Q_{h}}=\frac{T_{c}\Delta S_{\mathrm{eq},c}}{T_{h}\Delta S_{\mathrm{eq},h}}=\frac{T_{c}\left(T_{h}-T_{p}\right)}{T_{h}\left(T_{p}-T_{c}\right)}.\label{eq:copre}
\end{equation}

\section{MAIN RESULTS}

\subsubsection{The Trade-off Relation}

First we calculate the difference between $\varepsilon_{\mathrm{r}}$
and $\varepsilon$

\begin{equation}
\varepsilon_{\mathrm{r}}-\varepsilon=\frac{T_{c}\left(T_{h}-T_{p}\right)}{T_{h}\left(T_{p}-T_{c}\right)}-\frac{Q_{c}}{Q_{h}}=\frac{Q_{h}T_{c}\left(T_{h}-T_{p}\right)-Q_{c}T_{h}\left(T_{p}-T_{c}\right)}{Q_{h}T_{h}\left(T_{p}-T_{c}\right)}=\frac{-\beta_{h}Q_{h}-\beta_{c}Q_{c}+\beta_{p}(Q_{c}+Q_{h})}{(\beta_{c}-\beta_{p})Q_{h}}.\label{eq:dfferphi}
\end{equation}

The entropy change of environment is expressed as:
\begin{equation}
\Delta S_{\mathrm{en}}=-\beta_{h}Q_{h}-\beta_{c}Q_{c}-\beta_{p}Q_{p},\label{eq:Sen}
\end{equation}
Notice that $Q_{c}+Q_{h}+Q_{p}=0$, Eq. (\ref{eq:dfferphi}) can be
rewritten as

\begin{equation}
\varepsilon_{\mathrm{r}}-\varepsilon=\frac{\Delta S_{\mathrm{en}}}{(\beta_{c}-\beta_{p})Q_{h}}.
\end{equation}
Dividing both sides of this equation by $\varepsilon\tau$, we can
obtain

\begin{equation}
R(\frac{\varepsilon_{r}}{\varepsilon}-1)=\frac{\Delta S_{\mathrm{en}}}{(\beta_{c}-\beta_{p})\tau}.\label{eq:tradeoff}
\end{equation}

For a thermodynamic cycle, the entropy change of the reservoirs is
non-negative, i.e., $\Delta S_{\mathrm{en}}\geq0$ and $\varepsilon_{\mathrm{r}}\geq\varepsilon$.
Both inequalities become equalities only in the case of a reversible
cycle. 

\subsubsection{The Bound Given by the Thermodynamic Length}

We now focus on the $\Delta S_{\mathrm{en}}$ and rewrite Eq. (\ref{eq:Sen})
as

\begin{equation}
\Delta S_{\mathrm{en}}=-\sum_{v}\beta_{v}Q_{v}=-\sum_{v}\left(\Delta S_{\mathrm{eq},v}+\Sigma_{v}/\tau_{v}\right)=-\sum_{v}\Sigma_{v}/\tau_{v},
\end{equation}
Here, we have taken into account that, for a thermodynamic cycle,
the reversible part of the entropy change is zero, i.e., $\sum_{v}\Delta S_{\mathrm{eq},v}=0$.
We now discuss the bound for $-\sum_{v}\Sigma_{v}/\tau_{v}$. Considering
Eq. (\ref{eq:Q1}), we obtain

\begin{eqnarray}
-\sum_{v}\Sigma_{v}/\tau_{v} & = & -\sum_{v}\beta_{v}\int_{0}^{\tau_{v}}dt\mathrm{Tr}\left[H_{v}(t)\frac{d}{dt}\left\{ \mathcal{L}_{v}^{-1}(t)\frac{d}{dt}\left[\rho_{\mathrm{eq},v}(t)\right]\right\} \right]\nonumber \\
 & = & \sum_{v}\beta_{v}\int_{0}^{\tau_{v}}dt\mathrm{T_{r}}\left[\frac{d}{dt}H_{v}(t)\mathcal{L}_{v}^{-1}(t)\frac{d}{dt}\left[\rho_{\mathrm{eq},v}(t)\right]\right]-\sum_{v}\beta_{v}\left.\mathrm{T_{r}}\left[H_{v}(t)\mathcal{L}_{v}^{-1}(t)\frac{d}{dt}\left[\rho_{\mathrm{eq},v}(t)\right]\right]\right|_{0}^{\tau_{v}}.\label{eq:bypart}
\end{eqnarray}
First, we focus on the second part of this result. Considering the
control protocol discussed in Sec. (\ref{sec:THE-CONTROL-PROTOCOL}),
to complete the cycle, the Hamiltonians before and after the adiabatic
processes satisfy the following relationships: $H_{h}(0)=\left(T_{h}/T_{c}\right)H_{c}\left(\tau_{c}\right)$,
$H_{p}(0)=\left(T_{p}/T_{h}\right)H_{h}\left(\tau_{h}\right)$ and
$H_{c}(0)=\left(T_{c}/T_{p}\right)H_{p}\left(\tau_{p}\right)$. In
addition, by considering Eq. (\ref{eq:Gibbs}) and incorporating the
relationship between the Hamiltonians before and after the quench,
we find that the instantaneous Gibbs state of the density matrix remains
unchanged across the process, e.g., $\rho_{\mathrm{eq},h}(0)=\rho_{\mathrm{eq},c}(\tau_{c})$.
At the same time, the entropy remains unchanged before and after the
quenching process. This implies that the system's density matrix $\rho(t)$
of the system does not change. Considering the definition in Eq. (\ref{eq:1stexpandrho}),
after an adiabatic process, $\mathcal{L}_{v}^{-1}(t)\left[\frac{d}{dt}\rho_{\mathrm{eq},v}(t)\right]=\rho(t)-\rho_{\mathrm{eq},v}(t)$
remains unchanged. Thus, we proceed to calculate the second part as
follows: 
\begin{eqnarray}
 &  & \sum_{v}\beta_{v}\left.\mathrm{T_{r}}\left[H_{v}(t)\mathcal{L}_{v}^{-1}(t)\frac{d}{dt}\left[\rho_{\mathrm{eq},v}(t)\right]\right]\right|_{0}^{\tau_{v}}\nonumber \\
 & = & \beta_{c}\left.\mathrm{T_{r}}\left[H_{c}(t)L_{c}(t)\right]\right|_{0}^{\tau_{c}}+\beta_{h}\left.\mathrm{T_{r}}\left[H_{h}(t)L_{h}(t)\right]\right|_{0}^{\tau_{h}}+\beta_{p}\left.\mathrm{T_{r}}\left[H_{p}(t)L_{p}(t)\right]\right|_{0}^{\tau_{p}}\nonumber \\
 & = & \beta_{c}\mathrm{T_{r}}\left[H_{c}(\tau_{c})L_{c}(\tau_{c})\right]-\beta_{c}\mathrm{T_{r}}\left[\left(\beta_{p}/\beta_{c}\right)H_{p}\left(\tau_{p}\right)L_{c}(0)\right]\nonumber \\
 & + & \beta_{h}\mathrm{T_{r}}\left[H_{h}(\tau_{h})L_{h}(\tau_{h})\right]-\beta_{h}\mathrm{T_{r}}\left[\left(\beta_{c}/\beta_{h}\right)H_{c}\left(\tau_{c}\right)L_{h}(0)\right]\nonumber \\
 & + & \beta_{p}\mathrm{T_{r}}\left[H_{p}(\tau_{p})L_{p}(\tau_{p})\right]-\beta_{p}\mathrm{T_{r}}\left[\left(\beta_{h}/\beta_{p}\right)H_{h}\left(\tau_{h}\right)L_{p}(0)\right]\nonumber \\
 & = & 0.\label{eq:0}
\end{eqnarray}

For simplicity, we rewrite $\mathcal{L}_{v}^{-1}(t)\left[\frac{d}{dt}\rho_{\mathrm{eq},v}(t)\right]$
as $L_{v}(t)$. Taking Eq. (\ref{eq:0}) into Eq. (\ref{eq:bypart}),
we can perform further calculations

\begin{eqnarray}
-\Sigma_{v} & = & \beta_{v}\tau_{v}\int_{0}^{\tau_{v}}dt\mathrm{T_{r}}\left[\frac{d}{dt}H_{v}(t)\mathcal{L}_{v}^{-1}(t)\frac{d}{dt}\left[\rho_{\mathrm{eq},v}(t)\right]\right]\nonumber \\
 & \geq & \beta_{v}\left(\int_{0}^{\tau_{v}}dt\sqrt{\mathrm{T_{r}}\left[\frac{d}{dt}H_{v}(t)\mathcal{L}_{v}^{-1}(t)\frac{d}{dt}\left[\rho_{\mathrm{eq},v}(t)\right]\right]}\right)^{2}\nonumber \\
 & = & \beta_{v}\left(\int_{0}^{1}\frac{dt}{\tau_{v}}\cdot\tau_{v}\sqrt{\mathrm{T_{r}}\left[\frac{\partial\tilde{H_{v}}(s)}{\partial s}\frac{1}{\tau_{v}^{2}}\tilde{\mathcal{L}}_{v}^{-1}(s)\frac{d}{ds}\tilde{\rho}_{\mathrm{eq},v}(s)\right]}\right)^{2}\nonumber \\
 & = & \beta_{v}\mathcal{\mathscr{L}}_{v}^{2}.\label{eq:bound}
\end{eqnarray}
In the second line, we apply the Cauchy--Schwarz inequality, and
by rescaling time the as Eq. (\ref{eq:tsrho}), we arrive at the fourth
line to obtain the upper bound. Here, $\mathcal{L}_{v}$ is the thermodynamic
length, which is defined as \cite{Mehboudi2022,Chen2021}

\begin{equation}
\mathcal{\mathscr{L}}_{v}=\int_{0}^{1}ds\sqrt{\mathrm{T_{r}}\left[\frac{\partial\tilde{H_{v}}(s)}{\partial s}\tilde{\mathcal{L}}_{v}^{-1}(s)\frac{d}{ds}\tilde{\rho}_{\mathrm{eq},v}(s)\right]}.\label{eq:tlength}
\end{equation}

Substituting Eq. (\ref{eq:bound}) into Eq. (\ref{eq:tradeoff}),
we get our main result

\begin{equation}
R(\frac{\varepsilon_{r}}{\varepsilon}-1)\geq\sum_{v}\frac{\beta_{v}\mathcal{\mathscr{L}}_{v}^{2}}{(\beta_{c}-\beta_{p})\tau_{v}\tau}.
\end{equation}

On the other hand, by reversing the cycle, we obtain a three-reservoir
heat pump cycle. A similar trade-off relation can be derived through
analogous calculations:

\begin{equation}
P(\frac{\psi_{r}}{\psi}-1)\geq\sum_{v}\frac{\beta_{v}\mathcal{\mathscr{L}}_{v}^{2}}{(\beta_{c}-\beta_{h})\tau_{v}\tau},
\end{equation}
where $\dot{Q}_{h}$ denotes the rate of heat absorption from the
high-temperature reservoir, given by:

\begin{equation}
\dot{Q}_{h}=\frac{Q_{h}}{\tau}=\frac{k_{B}T_{h}\left[\Delta S_{\mathrm{eq},h}+\Sigma_{h}/\tau_{h}\right]}{\tau}.\label{eq:power}
\end{equation}
 $\psi$ and $\psi_{r}$ represent the COP of the heat pump in the
irreversible and reversible cases, respectively. Their explicit expressions
are given by:

\begin{equation}
\psi=\frac{Q_{h}}{Q_{p}}=\frac{T_{h}\left[\Delta S_{\mathrm{eq},h}+\Sigma_{h}/\tau_{h}\right]}{T_{p}\left[\Delta S_{\mathrm{eq},p}+\Sigma_{h}/\tau_{p}\right]},
\end{equation}
and

\begin{equation}
\psi_{r}=\frac{Q_{h}}{Q_{p}}=\frac{T_{h}\Delta S_{\mathrm{eq},h}}{T_{p}\Delta S_{\mathrm{eq},p}}=\frac{T_{h}\left(T_{p}-T_{c}\right)}{T_{p}\left(T_{h}-T_{c}\right)}.
\end{equation}

\subsubsection{Reduce to Two-reservoir Model}

In the high-temperature limit ($T_{h}\rightarrow\infty$ ), our model
naturally reduces to a two-reservoir configuration by treating the
energy exchanged by the high-temperature reservoir as work and considering
the intermediate heat reservoir as the effective high-temperature
reservoir. The thermal relaxation time $\tau_{h}$ becomes negligible
for a quantum system interacting with a bosonic thermal bath, as the
dissipation rate $\gamma\approx\gamma_{0}\left(\omega_{v}(t)\right)^{\alpha}\mathrm{\mathit{k_{B}T}}/\hbar\omega$
grows linearly with temperature ($\gamma\propto T$). This behavior
is a direct consequence of the high density of thermal excitations
in the bath. This temperature dependence causes complete quantum decoherence,
with the system reaching a maximally mixed state before any coherent
dynamics can develop. Crucially, the model's reduction to two reservoirs
emerges from this limiting behavior, where the high-temperature bath
effectively acts as an instantaneous thermalizer while the other reservoirs
maintain the system's fundamental thermodynamic coupling, and the
energy from the high-temperature bath can be seen as work. \cite{Chen1989,Yan1989}. 

Based on energy conservation, the energy supplied by the high-temperature
reservoir is given by $Q_{h}\equiv W=-(Q_{c}+Q_{p})$. The definition
of COP can be redefined as:

\begin{equation}
\varepsilon=\frac{Q_{c}}{W}=-\frac{T_{c}\left[\Delta S_{\mathrm{eq},c}+\Sigma_{c}/\tau_{c}\right]}{T_{c}\left[\Delta S_{\mathrm{eq},c}+\Sigma_{c}/\tau_{c}\right]+T_{p}\left[\Delta S_{\mathrm{eq},p}+\Sigma_{p}/\tau_{p}\right]}.\label{eq:cop-2}
\end{equation}
Meanwhile, Eq. (\ref{eq:copre}) reduced to

\begin{equation}
\varepsilon_{\mathrm{r}}=\frac{T_{c}}{T_{p}-T_{c}},
\end{equation}
which is denoted by the Carnot COP. inally, our result retains the
same form by simply setting $v=p,c$ and $\tau=\tau_{c}+\tau_{p}$.

We can apply the same procedure to the heat pump, which can be reduced
to a Carnot heat engine. In this case, $P$ is the power of the engine
given by

\begin{equation}
P=\frac{W}{\tau}=-\frac{T_{c}\left[\Delta S_{\mathrm{eq},c}+\Sigma_{c}/\tau_{c}\right]+T_{p}\left[\Delta S_{\mathrm{eq},p}+\Sigma_{p}/\tau_{p}\right]}{\tau}.\label{eq:power-1}
\end{equation}
while the efficiency can be written as:

\begin{equation}
\eta=\frac{W}{Q_{p}}=\frac{T_{h}\left[\Delta S_{\mathrm{eq},h}+\Sigma_{h}/\tau_{h}\right]}{T_{p}\left[\Delta S_{\mathrm{eq},p}+\Sigma_{h}/\tau_{p}\right]},
\end{equation}
and

\begin{equation}
\eta_{r}=\frac{T_{p}-T_{c}}{T_{p}}=\eta_{c}.
\end{equation}

Then, a similar trade-off relation of the reduced model can be derived: 

\begin{equation}
P(\frac{\eta}{\eta_{c}}-1)\geq\sum_{v}\frac{\beta_{v}\mathcal{\mathscr{L}}_{v}^{2}}{(\beta_{c}-\beta_{h})\tau_{v}\tau},
\end{equation}
where $v=p,c$ and $\tau=\tau_{c}+\tau_{p}$, and it can be interpreted
as the bound of the power--efficiency trade-off relation for a Carnot
engine. It has been shown that, for slowly driven finite-time Carnot
cycles, the upper bound of the power--efficiency trade-off can be
characterized by the thermodynamic length. The main conclusions of
Ref.\cite{Pietzonka2018,Miller2021} are as follows:

\begin{equation}
P\frac{\eta}{\eta_{\mathrm{C}}-\eta}\frac{T_{c}}{\Delta_{P}}\leq\frac{1}{2},
\end{equation}
and

\begin{equation}
\eta\leq\frac{\eta_{C}}{1+2T_{c}P_{w}f\left(\left|\frac{P_{w}}{P_{\omega}}\right|\right)/\left(\Delta P_{w}-2\Delta\mathcal{I}_{w}\right)}:=\eta^{Q}.
\end{equation}

With further calculation, these expressions can be rewritten as:

\begin{equation}
P\left(\frac{\eta_{c}}{\eta}-1\right)\geqslant\frac{2P^{2}}{\beta_{c}\Delta P},
\end{equation}
and

\begin{equation}
P_{w}\left(\frac{\eta_{c}}{\eta}-1\right)\geqslant\frac{\left\langle \left\langle \delta\dot{H}_{\lambda},\delta\dot{X}_{\lambda}\right\rangle \right\rangle }{\beta_{c}}.
\end{equation}

This means that lower bounds on the trade-off relations for both classical
and quantum scenarios are established in Ref. \cite{Pietzonka2018,Miller2021},
while here we present a lower bound derived via the thermodynamic
length. 

\section{THE CONTROL PROTOCOL OF THE TWO-LEVEL SYSTEM}

In the main text, we use a two-level system (TLS) as a model to illustrate
our results. The control protocol consists of three heat exchange
steps, during which the TLS is brought into thermal contact with a
reservoir $v(v=h,c$, or $p)$, and three adiabatic steps characterized
by sudden quenches in the system's Hamiltonian. The working substance
is a TLS with time-dependent Hamiltonian $H(t)=\hbar\omega_{v}(t)\sigma_{z}/2$,
where $\omega_{v}(t)$ is the energy splitting at time $t,\sigma_{z}$
is the Pauli matrix in $z$ direction, and $\hbar$ is Planck's constant.
The detail of the control protocols of the quantum tricycle are designed
as follows \cite{Cavina2017}:

(1) Heat exchange with reservoir $c$ (from point A to B): When the
TLS is coupled to reservoir $c$ with low temperature $T_{c}$, the
time-modulated field drives the frequency of the TLS according to
the cosine function $\omega_{c}(t)=\delta_{c}\left[\cos\pi\left(t/\tau_{c}\right)+\zeta_{c}\right]$
in the time interval $t\in$ $\left[0,\tau_{c}\right]$, where $\delta_{c}$
and $\zeta_{c}$ denote the parameters of the amplitude and the displacement,
respectively.

(2) Adiabatic expansion (from point B to C): The system is isolated
from any reservoir and the frequency suddenly shifted from $\omega_{c}\left(\tau_{c}\right)$
to $\omega_{h}(0)=T_{h}/T_{c}\omega_{c}\left(\tau_{c}\right)$. The
scaling factor $T_{h}/T_{c}$ is introduced, ensuring that the system
continuously remains in equilibrium state during a reversible quantum
tricycle. In addition, the rapidly changing condition of adiabatic
process prevents the TLS from adapting its configuration during the
process, and thus the probability density remains unchanged.

(3) Heat exchange with reservoir $h$ (from point C to $\mathrm{D})$
: The system is coupled to reservoir $h$ with high temperature $T_{h}$
in the time interval $t\in\left[0,\tau_{h}\right]$. The Hamiltonian
$H(t)$ of the system is concurrently modified by adjusting the frequency
according to the function $\omega_{h}(t)=\delta_{h}\left[\cos\pi\left(t/\tau_{h}\right)+\zeta_{h}\right]$,
where $\delta_{h}$ and $\zeta_{h}$ signify the parameters of the
amplitude and the displacement, respectively.

(4) First adiabatic compression (from point D to E): The system is
decoupled from the reservoir and a sudden quench is performed, in
which the frequency is changed from $\omega_{h}\left(\tau_{h}\right)$
into $\omega_{p}(0)=\left(T_{p}/T_{h}\right)\omega_{h}\left(\tau_{h}\right)$.

(5) Heat exchange with reservoir $p$ (from point E to F): An irreversible
heat exchange process follows by coupling the system with reservoir
$p$ at intermediate temperature $T_{p}$. The frequency of the system
is slowly changed in accordance with the function $\omega_{p}(t)=$
$\delta_{p}\left(\cos\pi\left(1-t/\tau_{p}\right)+\zeta_{p}\right)$
in a time interval $t\in\left[0,\tau_{p}\right]$. The parameters
$\delta_{p}$ and $\zeta_{p}$ are the amplitude and the displacement
in this process, respectively.

(6) Second adiabatic compression (from point F to A): Finally, the
system is disconnected from the reservoir and a sudden quench restores
the frequency back to $\omega_{c}(0)=$ $\left(T_{c}/T_{p}\right)\omega_{p}\left(\tau_{p}\right)$.

It is worth noting that, under this setup, the system adheres to the
quantum adiabatic theorem. In conjunction with Sec. (\ref{sec:THE-CONTROL-PROTOCOL}),
we consider the effective temperature $T_{\mathrm{eff},v}$ , which
can be calculated in the energy eigenbasis. It is given by $T_{\mathrm{eff},v}=\frac{\hbar\omega_{v}(t)}{k_{B}}\left(\ln\frac{\rho_{0}}{\rho_{1}}\right)^{-1}$,
where $\rho_{0}$ and $\rho_{1}$ denote the occupation probabilities
of the ground and excited states, respectively.

\subsubsection{The Relationships Between the Amplitude and the Displacement}

The choice of the parameters, amplitude $\delta_{v}$ and displacement
$\zeta_{v}$, in the modulation of the frequency $\omega_{v}(t)$
during adiabatic operations enables the system to transition between
equilibrium states corresponding to different reservoirs. In the context
of the quantum tricycle, it is essential that the frequencies of the
TLS at the end of one heat exchange process and at the beginning of
the subsequent heat exchange process are directly proportional. Specifically,
$\omega_{c}\left(\tau_{c}\right)=\left(T_{c}/T_{h}\right)\omega_{h}(0)$,
$\omega_{h}\left(\tau_{h}\right)=\left(T_{h}/T_{p}\right)\omega_{p}(0)$,
and $\omega_{p}\left(\tau_{p}\right)=\left(T_{p}/T_{c}\right)\omega_{c}(0)$.
These requirements lead to the following relationships: 
\begin{equation}
\frac{\delta_{c}\left(\zeta_{c}-1\right)}{\delta_{h}\left(\zeta_{h}+1\right)}=\frac{T_{c}}{T_{h}},\label{eq:deltar1}
\end{equation}
 
\begin{equation}
\frac{\delta_{h}\left(\zeta_{h}-1\right)}{\delta_{p}\left(\zeta_{p}-1\right)}=\frac{T_{h}}{T_{p}}\label{eq:deltar2}
\end{equation}
 and 
\begin{equation}
\frac{\delta_{p}\left(\zeta_{p}+1\right)}{\delta_{c}\left(\zeta_{c}+1\right)}=\frac{T_{p}}{T_{c}}.\label{eq:deltar3}
\end{equation}
By employing Eqs. (\ref{eq:deltar1})-(\ref{eq:deltar3}) in conjunction
with the given temperatures $T_{c},T_{h}$, and $T_{p}$, as well
as the displacement values of $\zeta_{c}$ and $\zeta_{h}$, and treating
the amplitude parameter $\delta_{c}$ as the independent variable,
we obtain the following concise relationships:

\[
\begin{gathered}\zeta_{p}=\frac{1+\zeta_{c}\zeta_{h}}{\zeta_{c}+\zeta_{h}}\end{gathered}
,
\]

\[
\delta_{h}=\frac{T_{h}\left(\zeta_{c}-1\right)}{T_{c}\left(1+\zeta_{h}\right)}\delta_{c},
\]
and
\[
\delta_{p}=\frac{T_{p}\left(\zeta_{c}+\zeta_{h}\right)}{T_{c}\left(1+\zeta_{h}\right)}\delta_{c}.
\]
In the subsequent discussion, we can investigate an arbitrary quantum
tricycle by varying the value of $\delta_{c}$. 

It is crucial to emphasize that the functioning of the cycle relies
on the condition $\sum_{v}Q_{v}^{0}\geq0$ , in accordance with the
principles of energy conversion. 

\subsubsection{The Dynamics of the Two-level System}

In our control protocols, the system’s evolution in the weak coupling
regime is described by the following Master Equation:

\begin{eqnarray}
\frac{d}{dt}\rho(t) & = & -\frac{i}{\hbar}[H(t),\rho(t)]\nonumber \\
 & + & \gamma_{v}(t)\left(n_{v}(t)+1\right)\left[\sigma_{-}\rho(t)\sigma_{+}-\frac{1}{2}\left\{ \sigma_{+}\sigma_{-},\rho(t)\right\} \right]\label{eq:meq}\\
 & + & \gamma_{v}(t)n_{v}(t)\left[\sigma_{+}\rho(t)\sigma_{-}-\frac{1}{2}\left\{ \sigma_{-}\sigma_{+},\rho(t)\right\} \right].
\end{eqnarray}
By following the operator definitions given in the main text, the
state of the TLS is represented by the vectorized form of the density
matrix $\rho(t)=\left(\begin{array}{llll}
\rho_{11} & \rho_{10} & \rho_{01} & \rho_{00}\end{array}\right)^{\mathrm{T}}$, where $\rho_{ij}(t)=\langle i|\rho(t)|j\rangle(i,j=0$ or 1$)$
denote the matrix elements of the density matrix. Based on Eq. (\ref{eq:meq}),
the super-operator $\mathcal{L}_{v}(t)$ can be expressed in matrix
form as follows:

\begin{equation}
\mathcal{L}_{v}(t)=\left(\begin{array}{cccc}
-\gamma_{v}(t)\left(n_{v}(t)+1\right) & 0 & 0 & \gamma_{v}(t)n_{v}(t)\\
0 & -\gamma_{v}(t)\left(n_{v}(t)+\frac{1}{2}\right)-i\omega_{v}(t) & 0 & 0\\
0 & 0 & -\gamma_{v}(t)\left(n_{v}(t)+\frac{1}{2}\right)+i\omega_{v}(t) & 0\\
\gamma_{v}(t)\left(n_{v}(t)+1\right) & 0 & 0 & -\gamma_{v}(t)n_{v}(t)
\end{array}\right).
\end{equation}

At time $t$, the instantaneous equilibrium state can be simplified
as follows:

\begin{equation}
\rho_{\mathrm{eq},v}(t)=\left(\begin{array}{llll}
\frac{n_{v}(t)}{2n_{v}(t)+1} & 0 & 0 & \frac{n_{v}(t)+1}{2n_{v}(t)+1}\end{array}\right)^{\mathrm{T}}
\end{equation}

The Drazin inverse $\mathcal{L}_{v}^{-1}(t)$ of $\mathcal{L}_{v}(t)$
is expressed as

\begin{equation}
\mathcal{L}_{v}^{-1}(t)=\left(\begin{array}{cccc}
-\frac{n_{v}(t)+1}{\gamma_{v}(t)\left(2n_{v}(t)+1\right)^{2}} & 0 & 0 & \frac{n_{v}(t)}{\gamma_{v}(t)\left(2n_{v}(t)+1\right)^{2}}\\
0 & \frac{1}{-\gamma_{v}(t)\left(n_{v}(t)+\frac{1}{2}\right)-i\omega_{v}(t)} & 0 & 0\\
0 & 0 & \frac{1}{-\gamma_{v}(t)\left(n_{v}(t)+\frac{1}{2}\right)+i\omega_{v}(t)} & 0\\
\frac{n_{v}(t)+1}{\gamma_{v}(t)\left(2n_{v}(t)+1\right)^{2}} & 0 & 0 & -\frac{n_{v}(t)}{\gamma_{v}(t)\left(2n_{v}(t)+1\right)^{2}}
\end{array}\right)
\end{equation}

By introducing the rescaled time parameter $s=t/\tau_{v}$ and and
applying the zeroth-order approximation from Eq. (\ref{eq:Q0}), the
absorbed heat $Q_{v}^{0}$ from bath $v$ an be estimated in the quasi-static
limit. Furthermore, the first-order correction of heat $Q_{v}^{1}$,
arising from the irreversible finite-time process, can be calculated
using Eq. (\ref{eq:Q1}), i.e.,

\begin{equation}
Q_{v}^{1}=\frac{\hbar}{2\tau_{v}}\int_{0}^{1}\tilde{\omega}_{v}(s)\mathrm{Tr}\left\{ \sigma_{z}\frac{d}{ds}\left\{ \tilde{\mathcal{L}}_{v}^{-1}(s)\frac{d}{ds}\left[\tilde{\rho}_{\mathrm{eq},v}(s)\right]\right\} \right\} ds
\end{equation}

\subsubsection{Optimization of Time Allocation Using the Lagrangian Method}

In the main text, we apply optimal control using the Lagrangian method.
Specifically, we consider the optimal configuration of the cycle that
maximizes the cooling rate $R$ under a given COP $\psi$. By employing
the Lagrangian method, we introduce the following Lagrangian function:

\begin{equation}
L\left(\tau_{c},\tau_{h},\tau_{p}\right)=R+\lambda_{1}\psi+\lambda_{2}\left(Q_{c}+Q_{h}+Q_{p}\right)\label{eq:Lagra}
\end{equation}
where $\lambda_{1}$ and $\lambda_{2}$ are the Lagrange multipliers
associated with the given COP and the law of energy conservation,
respectively. Form the Euler-Lagrange equations, we obtain

\begin{equation}
\frac{\partial L\left(\tau_{c},\tau_{h},\tau_{p}\right)}{\partial\tau_{c}}=0\label{eq:L1}
\end{equation}

\begin{equation}
\frac{\partial L\left(\tau_{c},\tau_{h},\tau_{p}\right)}{\partial\tau_{h}}=0\label{eq:L2}
\end{equation}
and

\begin{equation}
\frac{\partial L\left(\tau_{c},\tau_{h},\tau_{p}\right)}{\partial\tau_{p}}=0\label{eq:L3}
\end{equation}

By applying Eqs. (\ref{eq:L1}-\ref{eq:L3}) and substituting Eqs.
(\ref{eq:Qfin}), (\ref{eq:cop}) and (\ref{eq:cr}) into Eq. (\ref{eq:Lagra}),
we obtain the following time constraint equation after eliminating
the Lagrange multipliers:

\begin{equation}
\Delta S_{\mathrm{eq},h}\frac{\tau_{h}^{2}}{\Sigma_{h}}+\Delta S_{\mathrm{eq},p}\frac{\tau_{p}^{2}}{\Sigma_{p}}+\Delta S_{\mathrm{eq},c}\frac{\tau_{c}^{2}}{\Sigma_{c}}+2\left(\tau_{c}+\tau_{h}+\tau_{p}\right)=0.
\end{equation}

\subsubsection{Numerical results}

\begin{figure}
\begin{centering}
\includegraphics[scale=0.3]{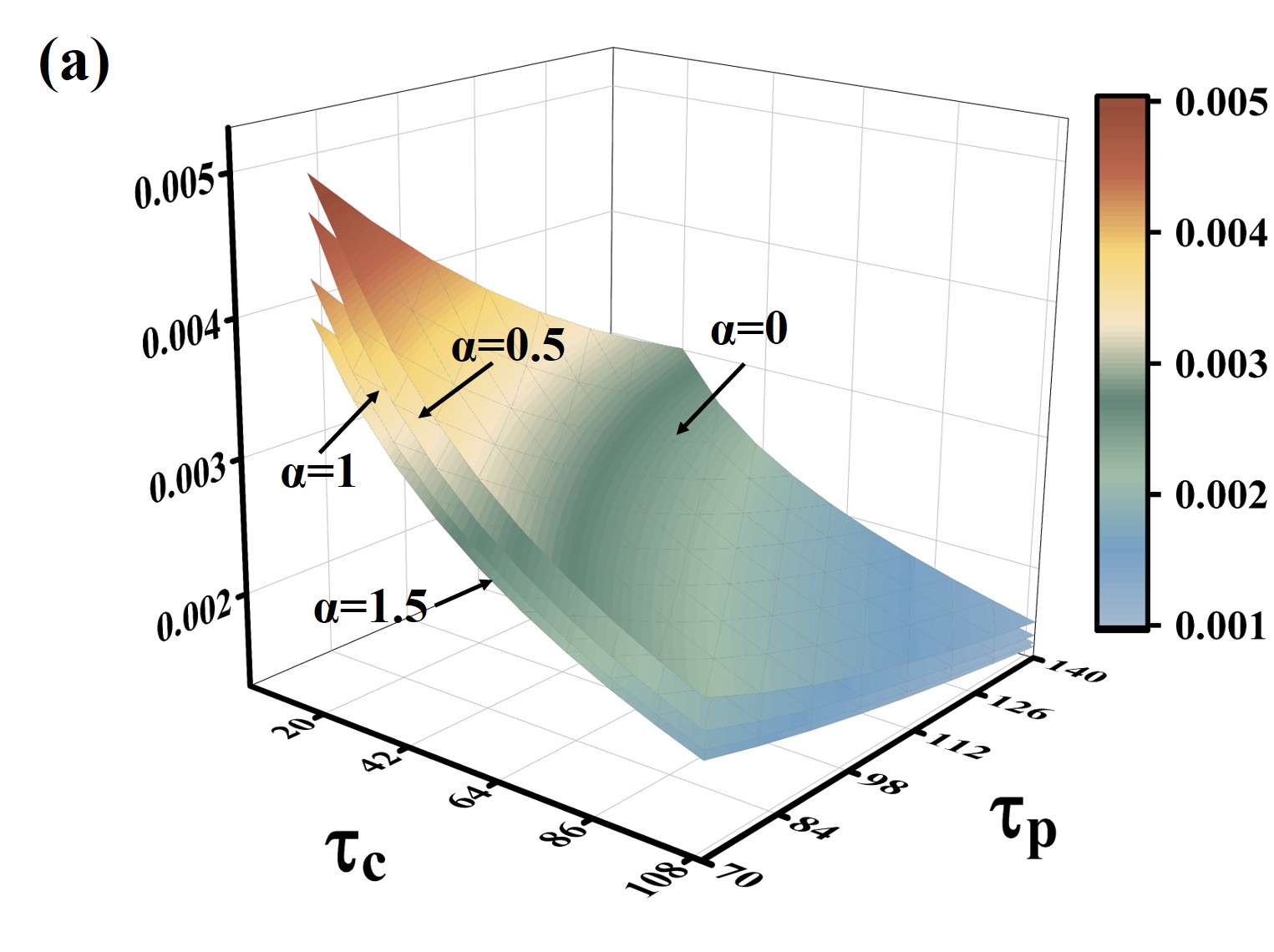}\includegraphics[scale=0.3]{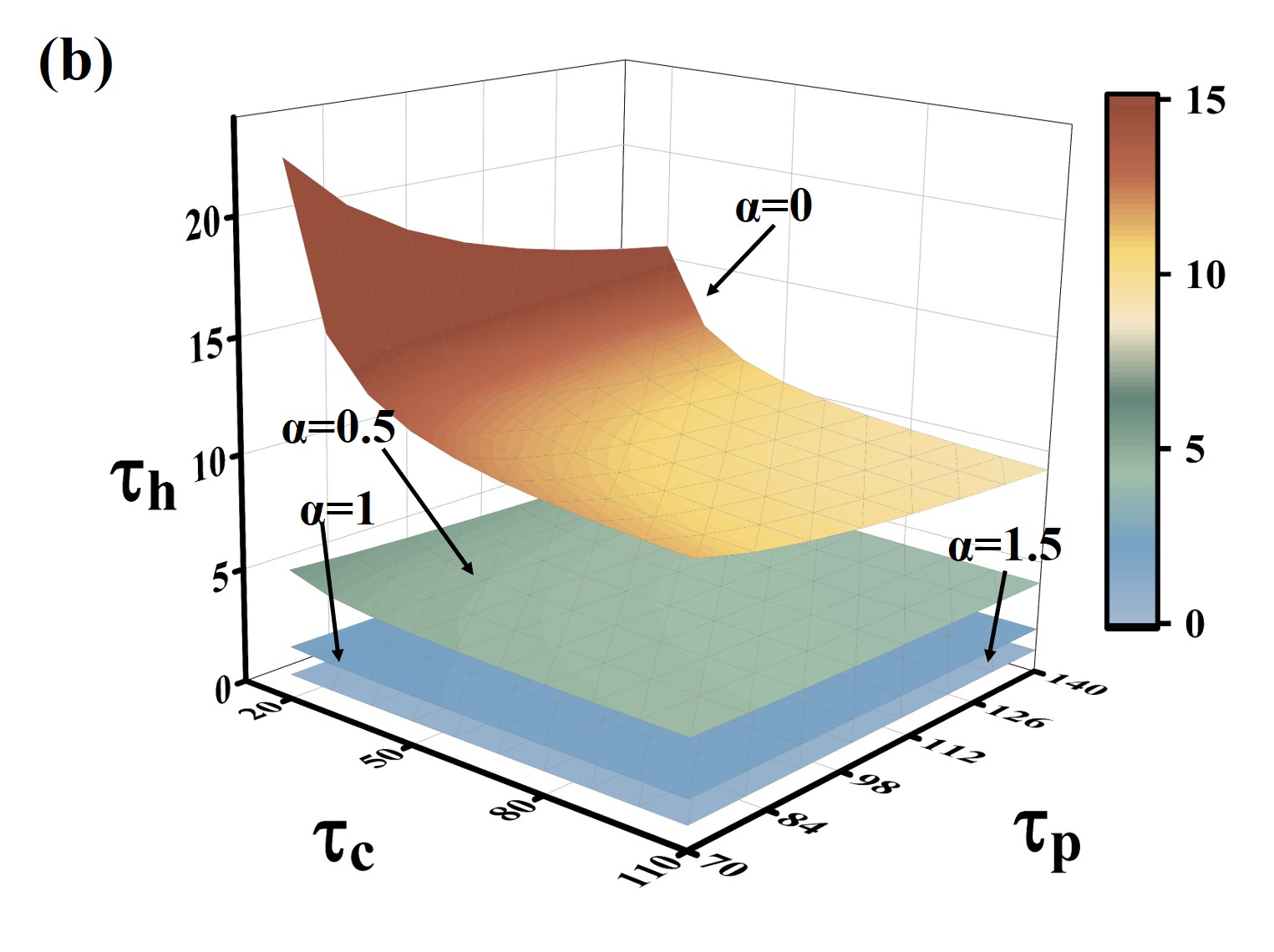}
\par\end{centering}
\caption{This figure presents three-dimensional visualizations of key performance
metrics. Fig. (a) shows the difference between the left-hand side
(LH) and right-hand side (RH) of the thermodynamic bound, defined
as $(\mathrm{LH}=R\left(\varepsilon_{r}/\varepsilon-1\right))$ and
$(\mathrm{RH}=\bar{\mathcal{\mathscr{L}}}^{2}/\tau)$, plotted against
the contact time with cold reservoir $c$ $(\tau_{c})$ and auxiliary
time with reservoir $p$ $(\tau_{p})$. Fig. (b) illustrates the variation
of the optimized interaction time $(\tau_{h})$ with respect to $(\tau_{c})$
and $(\tau_{p})$. Each surface corresponds to a different value of
the spectral parameter $(\alpha)$, while all other parameters are
fixed as $\gamma_{0}=k_{B}=\hbar=1$, $\delta_{c}=k_{B}T_{c}$, $\zeta_{c}=\zeta_{h}=2$,
$T_{h}=6$, $T_{p}=2.4$, and $T_{c}=2$ with $\varepsilon_{\mathrm{r}}=3$.}
\end{figure}

Here, we present additional numerical verification. Figure\,2 shows
our numerical results for different values of $\alpha$, corresponding
to different Ohmic spectra. The validation of our inequality is depicted
in Fig.\,2(a), where we plot the three-dimensional surface of the
difference between $\mathrm{LH}=R\left(\varepsilon_{r}/\varepsilon-1\right)$
and $\mathrm{RH}=\bar{\mathcal{\mathscr{L}}^{2}}/\tau$ as functions
of $\tau_{c}$ and $\tau_{p}$. It is observed that all the surfaces
lie above zero, confirming the validity of our inequality. Moreover,
smaller surface values indicate a tighter bound. As $\alpha$ increases,
the overall magnitude of the surface decreases, implying that the
upper bound becomes tighter. Additionally, for each value of $\alpha$,
the bound improves further as $\tau_{c}$ and $\tau_{p}$ increase.

Figure\,2(b) illustrates the surface of $\tau_{h}$ as a function
of $\tau_{c}$ and $\tau_{p}$ for different values of $\alpha$.
It can be observed that the contact time $\tau_{h}$ with the high-temperature
reservoir decreases as $\tau_{c}$ and $\tau_{p}$ increase. Moreover,
for a fixed pair $\left(\tau_{c},\tau_{p}\right)$, increasing $\alpha$
leads to a reduction in the overall value of $\tau_{h}$. This behavior
arises because a larger $\alpha$ enhances the transition rate $\gamma$,
thereby accelerating the thermalization process. At the same time,
within our setup, if $\alpha$ increases beyond a certain point, $\tau_{h}$
falls below 1, violating the assumptions of the slowly driven model.
Therefore, we restrict our verification to select cases with $\alpha$
ranging from 0 to 1.2 to ensure the validity of our results across
different Ohmic spectra.

\section*{ACKNOWLEDGMENTS}

This work has been supported by the Natural Science Foundation of
Fujian Province (2023J01006), National Natural Science Foundation
of China (12364008 and 12365006), and Fundamental Research Fund for
the Central Universities (20720240145).\\